\documentclass[12pt]{article}
\usepackage{hyperref}

\usepackage{epsfig}
\usepackage{graphicx}
\usepackage{a4}
\usepackage{latexsym}
\usepackage{times}

\graphicspath{{Pics/}}

\usepackage[normalem]{ulem} 
\usepackage {ulem} 
\usepackage{colordvi}

\usepackage{pslatex}
\usepackage[latin1]{inputenc}
\usepackage[T1]{fontenc}
\usepackage{txfonts}

\def\I {\rm{i}}
\def\MSbar{\overline{\mathrm{MS}}}
\def\calO{{\cal O}}

\def\lifo {{\rm{Li_4}}}
\def\lifi {{\rm{Li_5}}}
\def\ep   {\epsilon}

\def\EulerGamma {\gamma_E}

\def\b#1{\beta_#1}
\def\z#1{{\zeta_{#1}}}
\def\a4{{\rm{Li_4({1/2})}}}

\def\lntwo{{\ln 2}}
\def\lntwos{{\ln^2 2}}
\def\lntwoc{{\ln^3 2}}
\def\lntwof{{\ln^4 2}}
\def\lntwoq{{\ln^5 2}}

\def\lnbeta{\ln \beta}
\def\lnbetas{\ln^{\,2} \beta}

\def\ca{{C^{}_A}}
\def\tr{{T^{}_R}}
\def\cf{{C^{}_F}}
\def\nf{{n^{}_{\! f}}}
\def\nh{{n^{}_{\! h}}}
\def\nl{{n^{}_{\! l}}}

\def\asnf{\alpha_{s}^{(\nf)}}
\def\x#1{{x_{#1}}}

\begin{document}

\begin{titlepage}
\noindent
DESY 09-064 \\
HEPTOOLS 09-016 \\
SFB/CPP-09-35 \\
YITP-SB-09-08 \\
May 2009
\vspace{1.0cm}
\begin{center}
\LARGE {\bf
The QCD form factor of heavy quarks at NNLO 
}\\
\vspace{1.9cm}
\large
J. Gluza$^{a}$, A. Mitov$^{b}$, S. Moch$^{c}$ and T. Riemann$^{c}$ \\
\vspace{1.2cm}
\normalsize
{\it
$^{a}$Department of Field Theory and Particle Physics,
Institute of Physics \\[0.5ex]
University of Silesia, Uniwersytecka 4, PL-40007 Katowice, Poland \\[.5cm]
$^{b}$C. N. Yang Institute for Theoretical Physics \\[0.5ex]
Stony Brook University, Stony Brook, NY 11794, USA\\[.5cm]
$^{c}$Deutsches Elektronensynchrotron DESY \\[0.5ex]
Platanenallee 6, D--15738 Zeuthen, Germany}
\vfill
\large {\bf Abstract}
\vspace{-0.2cm}
\end{center}
We present an analytical calculation of the two-loop QCD corrections 
to the electromagnetic form factor of heavy quarks. 
The two-loop contributions to the form factor are reduced to linear combinations of master integrals, 
which are computed through higher orders in the parameter of 
dimensional regularization $\epsilon=(4-D)/2$.
Our result includes all terms of order $\epsilon$ at two loops 
and extends the previous literature.
We apply the exponentiation of the heavy-quark form factor 
to derive new improved three-loop expansions in the high-energy limit.
We also discuss the implications for predictions of massive $n$-parton
amplitudes based on massless results in the limit, where the quark mass is small
compared to all kinematical invariants.
\\
\vspace{1.0cm}
\end{titlepage}

\newpage

%
%
\section{Introduction}
\label{sec:intro}
%
%

The electromagnetic form factor of quarks comprises the simplest example
of a scattering amplitude in Quantum Chromodynamics (QCD) 
and is of basic importance both for theory and phenomenology.
To mention one example, the heavy-quark form factor is directly linked 
to phenomenological predictions for the forward-backward asymmetry 
for inclusive bottom quark production in electron-positron annihilation.
In a wider theoretical context, the heavy-quark form factor is a quantity of interest in its own.
Being gauge invariant, it provides the simplest instance 
for the study of mass effects in QCD hard scattering processes 
including higher-order quantum corrections and, 
as a key building block, it also relates 
to many other hard scattering processes involving massive quarks.
For these reasons the form factor of massive quarks and in particular its QCD
radiative corrections have received much attention in recent years.

In an impressive series of papers the two-loop QCD corrections to the heavy-quark form factors
have been obtained for the vector- and axial-vector coupling as well as the 
anomaly contributions~\cite{Bernreuther:2004ih,Bernreuther:2004th,Bernreuther:2005rw}.
Besides being mandatory for precision phenomenology, these explicit results 
for the massive form factor also exhibit certain universal features 
of higher order radiative corrections. 
It has long been known, that the leading logarithms of 
Sudakov type~\cite{Sudakov:1954sw} in heavy-quark mass $m$ can be resummed~\cite{Collins:1980ih}, while the 
exponentiation of the complete collinear and infrared singularities 
of the massive form factor within dimensional regularization 
has been addressed only rather recently~\cite{Mitov:2006xs}.

As an immediate consequence it is now possible to derive important partial information 
about massive $n$-parton gauge amplitudes from purely massless
calculations at any order in perturbation theory~\cite{Mitov:2006xs}, see also~\cite{Catani:2000ef,Becher:2007cu}.
This exploits the factorization properties in the soft and collinear limit 
of the respective amplitudes, i.e. the one with all partons massless~\cite{Catani:1998bh,Sterman:2002qn} 
and the amplitude with, say, fermions of mass $m$ in the presence of a typical large
kinematical invariant $Q^2$ in the limit $m^2/Q^2 \to 0$.
The procedure results in a universal and process independent 
multiplicative relation between the two amplitudes.
The concept has been put to test in various applications up to two loops, e.g.
the derivation of the QCD corrections to hadro-production of top-quarks~\cite{Czakon:2007ej,Czakon:2007wk} 
or the Quantum Electrodynamics (QED) correction to Bhabha scattering~\cite{Becher:2007cu}. 
Currently, this is an active area of research~\cite{Mitov:2009sv,Becher:2009kw}.

In the present paper we study the two-loop QCD corrections for the vector form
factor of heavy quarks in dimensional regularization. 
We first perform an independent check of Ref.~\cite{Bernreuther:2004ih}. 
Subsequently, we provide new results by extending the two-loop results 
in $D=4-2\epsilon$ dimensions up to terms of order $\epsilon$.
To that end the relevant master integrals at two loops have been computed 
to sufficiently high orders in $\epsilon$ and the results for those are given as well.
We are then in the position to discuss several applications.
After testing the $D$-dimensional exponentiation to order $\epsilon$ at two loops, 
we can utilize its predictive power to derive a new improved three-loop
prediction for the form factor to logarithmic accuracy in the heavy-quark mass $m$. 
In the same spirit, we extend the results for the universal $Z$ factor 
relating massless and massive amplitudes in the small-mass limit. 
Finally, as a by-product of our calculation, we also investigate the threshold 
limit of the heavy-quark form factor up to two loops and we discuss the Coulomb
singularities for color-singlet and octet currents.

The outline of the paper is as follows. 
In Section~\ref{sec:calculation} we recall the definition of the vector form factor 
of massive quarks and sketch the important parts of our computation.
We also provide some details of the integral reduction and the relevant master integrals.
Our main result, the two-loop QCD contributions for the form factor up to order $\epsilon$, 
is presented in Sec.~\ref{sec:results} along with expansions in kinematical limits 
(threshold and high energy production).
Finally, Sec.~\ref{sec:appl} discusses the above mentioned applications, 
i.e. the exponentiation which leads to a new three-loop prediction 
for the form factor and to improved formulae for relating massless and massive
amplitudes through three loops.
We conclude in Sec.~\ref{sec:conclusion}. 

%
%
%
\section{Calculation}
\label{sec:calculation}
%
%

The coupling of a vector current to a heavy-quark pair can be conveniently
written in terms of a vertex function $\Gamma_\mu$. 
Given a photon of virtuality $Q^2$ (we take space-like $q^2 = - Q^2 < 0$ throughout this Section) 
the expression for $\Gamma_\mu$ reads
\begin{equation}
\label{eq:ffdef}
\Gamma_\mu(k_1,k_2)  \: = \:
{\rm i} e_{\rm q}\,
{\bar \psi}(k_1)\,
\left(
\gamma_{\mu\,}\, {\cal F}_1 (Q^2,m^2,\alpha_{s})
-
{{\rm i} \over 2 m}
\,
\sigma_{\mu \nu\,} q^\nu \, {\cal F}_2 (Q^2,m^2,\alpha_{s}) \right) \psi(k_2)
\, .
\end{equation}
Here the external quark (anti-quark) of incoming momentum $k_1$ ($k_2$)
is on-shell with $m$ denoting its mass and $e_{\rm q}$ its
charge, thus $k_1^2 = m^2$ (and $k_2^2 = m^2$), and it is $q+k_1+k_2=0$.
The scalar functions ${\cal F}_1$ and ${\cal F}_2$
on the right-hand side are the electric and magnetic 
space-like quark form factors, i.e. our main quantities of interest.

%
\begin{figure}[ht]{
  \centering
  {
    \scalebox{0.6}{\includegraphics[clip]{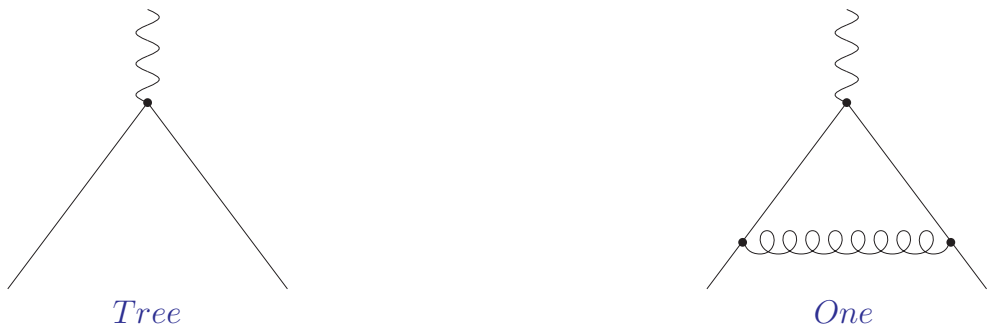}}
  }
  \caption{\small
  \label{fig:massiveFF1}
  Feynman diagrams contributing to the quark form factor of massive quarks up
  to one loop. 
}
}
\end{figure}
\begin{figure}[ht]{
  \centering
  {
    \scalebox{0.7}{\includegraphics[clip]{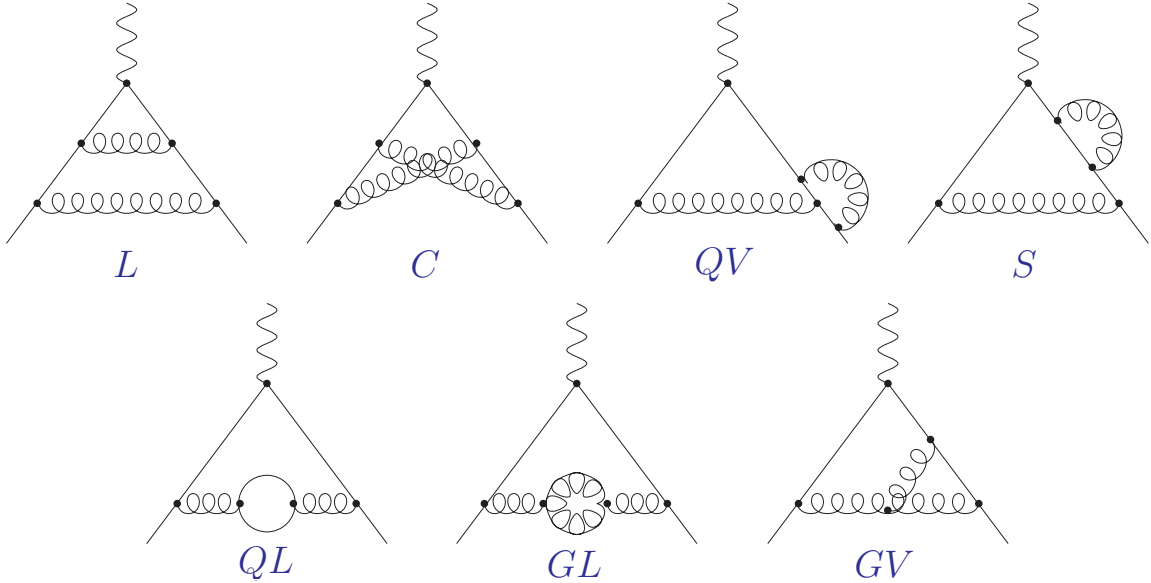}}
  }
  \caption{\small
  \label{fig:massiveFF2}
  Feynman diagrams contributing to the quark form factor of massive quarks at
  two loops. The diagrams $GV$, $QV$ and $S$ carry a symmetry factor of $2$. 
  Diagram $QL$ denotes both heavy and light quarks in the loop and 
  diagram $GL$ represents the gluon and the ghost loop contribution 
  to the gluon self-energy.
}
}
\end{figure}
%
With the help of suitable projections of $\Gamma_\mu$ (see e.g. Ref.~\cite{Bernreuther:2004ih}) 
${\cal F}_1$ and ${\cal F}_2$ in Eq.~(\ref{eq:ffdef}) can be computed.
They are gauge invariant, but in general, at higher orders in perturbative QCD divergent.
Both, ${\cal F}_1$ and ${\cal F}_2$ enjoy a power expansion in the strong coupling $\alpha_{s}$ 
at the scale $\mu^2$ (to be employed also for all other quantities throughout this article) 
\begin{eqnarray}
\label{eq:aexp}
  {\cal F}(\alpha_{s}) 
&=&  \sum\limits_{i=0}^{\infty} 
 \left(
{\alpha_{s}(\mu^2) \over 4\pi}
\right)^i
\, {\cal F}^{(i)}
\, \equiv \, \sum\limits_{i=0}^{\infty} \left(a_s(\mu^2)\right)^{\,i}\, {\cal F}^{(i)}\, ,
\end{eqnarray}
where we have introduced the shorthand notation $a_s \equiv \alpha_{s} / (4\pi)$.
All necessary Feynman diagrams 
entering the computation of the electric and magnetic form factors ${\cal F}_1$ and ${\cal F}_2$ 
through NNLO in QCD are displayed in Figs.~\ref{fig:massiveFF1} and~\ref{fig:massiveFF2}.
At the two loop level, we have summarized the contributions with a self-energy correction 
to the gluon line into the two diagrams $QL$ and $GL$ as follows.
Diagram $QL$ represents both the heavy and the $\nl$ light quarks in the loop correction to the gluon propagator. 
Likewise diagram $GL$ stands for the gluon and the ghost loop. 
Throughout this article, $\nf=\nl+1$ is the total number of flavors, which is the sum of
one heavy ($\nh=1$) and $\nl$ light quarks. 
The notation in Fig.~\ref{fig:massiveFF2} is taken over from the corresponding
calculation for the form factor of massless quarks, see e.g. Ref.~\cite{Moch:2005id}.

\subsection{Master integrals}
The number of Feynman diagrams in Figs.~\ref{fig:massiveFF1} and~\ref{fig:massiveFF2} is relatively small.
Thus, their treatment and the reduction of the corresponding Feynman integrals proceeds in a standard way.
The reduction has been achieved with the Laporta-Remiddi algorithm~\cite{Laporta:1996mq,Laporta:2001dd}, 
as implemented in the {\ttfamily Maple} program {\ttfamily AIR}~\cite{Anastasiou:2004vj}. 
In this way a list of so-called master integrals, possibly with additional
irreducible scalar products of loop momenta in the numerator (in short: `numerators') 
has been obtained 
together with the corresponding algebraic relations between the master integrals 
and the form factors ${\cal F}_1$ and ${\cal F}_2$.

\begin{figure}[htb]{
  \centering
  {
    \scalebox{0.5}{\includegraphics[clip]{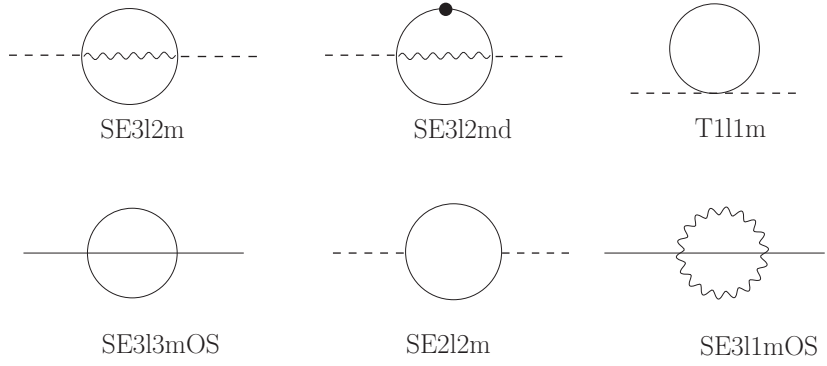}}
  }
  \caption{\small
  \label{fig:MIsSE}
  The tadpole and the two-point master integrals. The suffix OS denotes on-mass-shell master integrals.
}
}
\end{figure}

\begin{figure}[htb]{
  \centering
  {
    \scalebox{0.6}{\includegraphics[clip]{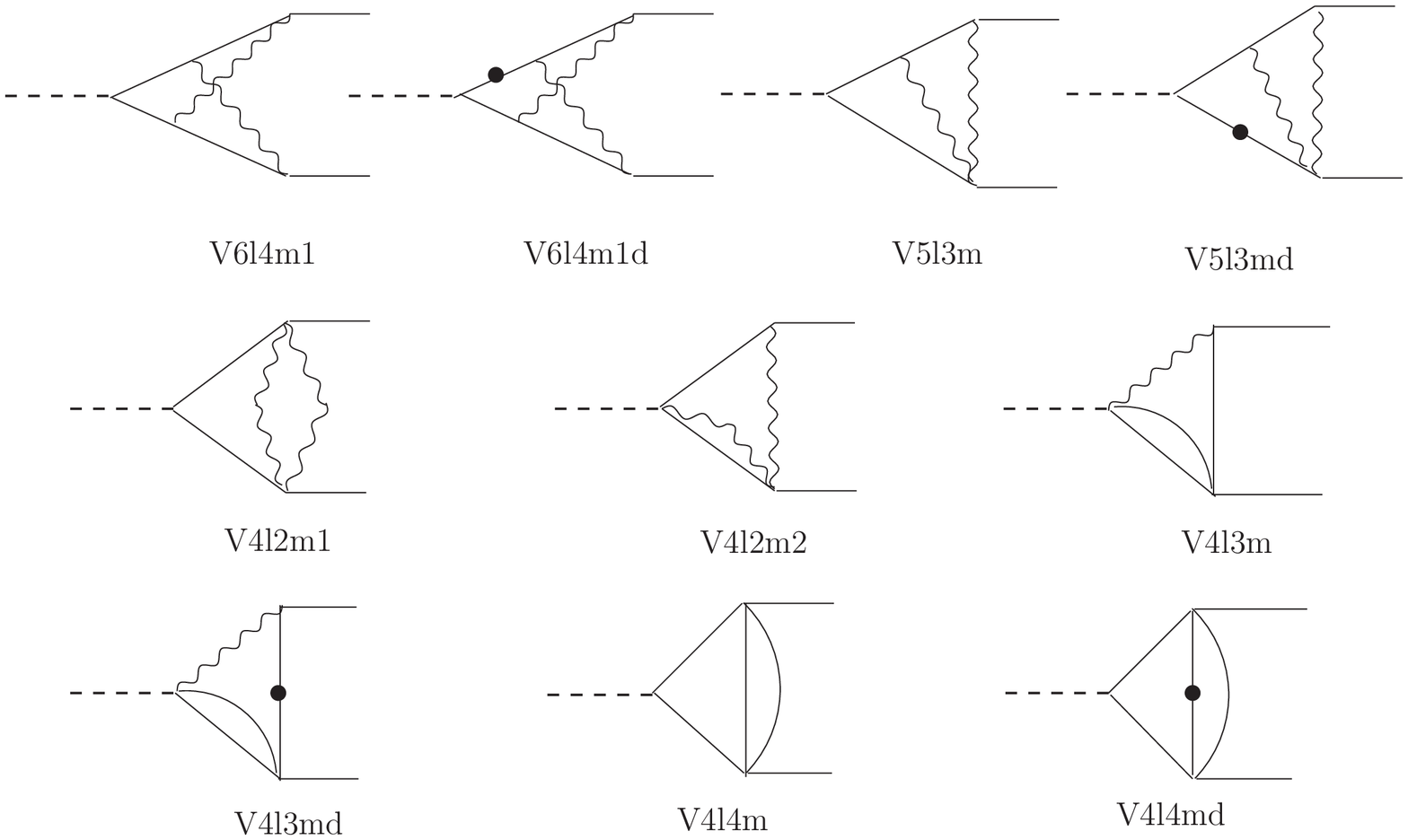}}
  }
  \caption{\small
  \label{fig:MIsV}
  The three-point master integrals. 
}
}
\end{figure}

We display all master integrals in Figs.~\ref{fig:MIsSE} and~\ref{fig:MIsV} 
and tabulate them in Tab.~\ref{tab:MIs}. 
We found it convenient to replace master integrals with numerators by so-called dotted ones.
Dotted master integrals are free of numerators but have higher powers of propagators. 
They are independent of the (arbitrary) momentum configurations inside the loop graphs.
This transformation can be achieved with appropriate integration-by-parts (IBP)
relations.

%
%
The Feynman integrals are defined as follows:
\begin{eqnarray}
  \label{eq:def-fi}
  {\cal  F}_L(X) &=& \frac{e^{L\ep\gamma_E}}{\left(i\pi^{d/2}\right)^L} \int \frac{d^d k_1 \ldots d^d k_L~~X(k_1\ldots k_L)}
  {(q_1^2-m_1^2)^{\nu_1} \ldots (q_j^2-m_j^2)^{\nu_j} \ldots
       (q_N^2-m_N^2)^{\nu_N}  }  ,
\end{eqnarray}
where $L$ is the number of loops and $X$ some numerator structure. With this definition, for example, the one-loop tadpole \texttt{T1l1m} becomes:
\begin{eqnarray}
\mathtt{T1l1m} 
&=& 
 \frac{     e^{\epsilon\gamma_E}}{i\pi^{D/2}} 
\int \frac{d^{D}k}{k^2-m^2} 
\nonumber\\
&=&
-\left(m^2 \right)^{1-\ep}~e^{\ep \EulerGamma} ~ \Gamma(-1 + \ep) 
\nonumber\\
&=&
\left(m^2 \right)^{1-\ep}
\left[ 
{1 \over \ep}
 +1
+\left(1+ {\z2 \over 2}
 \right) \* \ep 
+\left(1+ {\z2 \over 2}
 -        {\z3 \over 3}
\right) \* \ep^2 \right] 
+ ~{\cal O}(\ep^3).
\label{norm}
\end{eqnarray}
Actually, all master integrals have already been considered 
in a project concerned with the two-loop QED corrections to Bhabha scattering~\cite{Czakon:2004tg,Czakon:2004wm}, 
as well as during the previous two-loop QCD computation of the heavy-quark form factor in Ref.~\cite{Bernreuther:2004ih} 
(see also the References in the last column of Tab.~\ref{tab:MIs}). 
Because we are aiming at the determination of ${\cal F}_1$ and ${\cal F}_2$ including
the ${\cal O}(\epsilon)$ terms at two loops, the master integrals are needed here to higher powers in $\ep$.
Therefore we will point out specifically the required extensions (see Tab.~\ref{tab:MIs}).

For instance, the on-shell master integral {\ttfamily SE3l3mOS} (see Fig.~\ref{fig:MIsSE}) 
has to be evaluated until terms of the order  ${\cal O}(\epsilon^3)$. 
The solution for {\ttfamily SE3l3mOS}, which is, together with the other 
on-mass-shell master {\ttfamily SE3l1mOS}, actually one of the simplest masters,
has been obtained using the package {\ttfamily ON-SHELL2}~\cite{Fleischer:1999tu}, 
{\small
\begin{eqnarray}
  \label{bh-19a}
\mathtt{SE3l3mOS}
&=& 
\left( m^2\right)^{1-2\ep} 
  \Biggl\{ 
  {3\over 2\*\ep^2} + {17\over 4\*\ep} + {59\over8} + {3\* \z2\over2} 
+ 
\ep\*\biggl[ {65\over16} + {49\* \z2\over4} -  \z3 \biggr]
\\
& &\mbox{}
+
 \ep^2\* \biggl[ -{1117\over32} + 28\* \z3 + {21\* \z4\over8} 
+  \z2\*\biggl({475\over8} - 48\*\lntwo \biggr) -
   {17\* \z3\over6} \biggr]  
\nonumber \\
& &\mbox{}
+ \ep^3\* \biggl[ -{13783\over64} + 182\* \z3 - {1545\* \z4\over16} + 8\*\lntwof 
+
   192\*\lifo({1/2}) 
\nonumber \\
& &\mbox{}\hspace*{5mm}
+  \z2\*\biggl({3745\over16} - 312\*\lntwo 
+ 96\*\lntwos -  \z3\biggr) -
   {59\* \z3\over12} - {3\* \z5\over5} \biggr]
\nonumber \\
& &\mbox{}
+
 \ep^4\* \biggl[ -{114181\over128} + 805\* \z3 - {21219\* \z4\over32} - 930\* \z5 + {93\* \z6 \over32} +
   624\* \z4\*\lntwo + 52\*\lntwof 
\nonumber \\
& &\mbox{}\hspace*{5mm}
- {48\*\lntwoq \over 5} 
+ 1248\*\lifo({1/2}) +
   1152\*\lifi({1/2}) 
+  \z2\*\biggl({26563\over32} - 52\* \z3 - 1380\*\lntwo 
\nonumber \\
& &\mbox{}\hspace*{5mm}
+ 624\*\lntwos -
     192\*\lntwoc - {49\* \z3\over6} \biggr)
- {65\* \z3\over24} + { \z3^2\over3} -
   {17\* \z5\over10} \biggr] 
   \Biggr\} 
\nonumber
+ ~{\cal O}(\ep^5).
\label{seOS}
\end{eqnarray}
}
The computation of QCD corrections to form factors is a one-scale problem with variable $Q^2/m^2$ and
the results can be expressed in terms of harmonic polylogarithms (HPLs) up to weight five 
$H_{m_1,...,m_w}(x)$, $m_j = 0,\pm 1$. 
The HPLs were introduced in~\cite{Remiddi:1999ew}. They depend on the conformal variable $x$,  
\begin{equation}
  \label{eq:spacelike-xdef}
  \displaystyle
  x = {\sqrt{Q^2 + 4m^2} - \sqrt{Q^2} \over \sqrt{Q^2 + 4m^2} + \sqrt{Q^2}} 
  \, ,
\end{equation}
with $Q^2 > 0$ and $0 \le x \le 1$.

%
\begin{table}[t]
  \centering
\begin{tabular}{|l|r|c|c|c|c|c|c|c|c|l|}
\hline
Masters & $\ep^n,~~ n$~~~&weight&                       L &  C  & QV &  S  &  QL & GL & GV  & $\ep^n$, $n$, in Ref.\\
\hline \hline
{\tt T1l1m}$^\ast$ &--1 $\cdots$ 5 & 6 &       + &  +  & +  &  +  & +   & --  &  +  & \cite{'tHooft:1979xw}\\
\hline
{\tt SE2l2m}$^\ast$    & --1 $\cdots$ 4 & 5 &   + &  +  & +  &  +  & +   & -- &  +  &  
 \cite{'tHooft:1979xw}\\
{\tt SE3l1mOS} & --2 $\cdots$ 3 & 5 &            + &  +  & +  &  +  &  -- &  + &  +  &    
$\ep^2$ {\cite{Fleischer:1999tu}}\\
{\tt SE3l2m[d]} &  --2 $\cdots$ 4 & 6[6] &         + &  +  & +  &  +  &  -- &  --&  +  &  
$\ep^2$ \cite{Fleischer:1999hp,Davydychev:2003mv}\\
{\tt SE3l3mOS} & --2 $\cdots$  3& 5 &            -- &  +  & +  &  -- &  +  &  --&  --  & 
$\ep^4$ {\cite{Broadhurst:1990ei,Broadhurst:1991fi,Fleischer:1999tu}} \\
\hline
{\tt V4l2m1} & --2 $\cdots$ 3& 5 &             + &   +  & -- &  -- &  -- &    + &  +   & 
$\ep^0$ \cite{Bonciani:2003te}\\
{\tt V4l2m2} & --2 $\cdots$ 2& 5 &            + &   +  & +  &  +  &  -- &  --  &  +   & 
$\ep^0$ \cite{Bonciani:2003te}\\
{\tt V4l3m[d]} & --2 $\cdots$ 2& 5[5] &          -- &   +  & +  &  -- &  -- &  --  & --  & $
\ep^0$ \cite{Bonciani:2003te}\\
{\tt V4l4m[d]} & --2 [0] $\cdots$  2 & 5[5] &         -- &   +  & +  &  -- &  +  &   -- & --  & $\ep^0$ \cite{Bonciani:2003te}\\
\hline
{\tt V5l3m[d]} & 0 [--1] $\cdots$ 1 & 5[4] &          + &   +  & -- &   -- & --  &  --  &  +   & $\ep^0$ \cite{Bonciani:2003te}\\
\hline
{\tt V6l4m1[d]} &  --1 $\cdots$ 1 & 5 &       -- &   +  & -- &   -- &  -- &  --  &  --  & 
$\ep^0$  \cite{Bonciani:2003te}\\
\hline
\hline
\end{tabular}
\vspace*{3mm}
\caption{\small
  \label{tab:MIs}
  The master integrals entering the two-loop vertex
  diagrams in Fig.~\ref{fig:massiveFF2}. 
  The notations follows Refs.~\cite{Czakon:2004tg,Czakon:2004wm}.
  Stars `$\ast$' denote one-loop master integrals, the suffix `OS' on-mass-shell
  functions, and the suffix `d' dotted masters.
}
\end{table}
One-scale master integrals can be naturally solved by the method of differential equations~\cite{Kotikov:1991hm,Remiddi:1997ny} 
which allows the determination of the master integrals analytically to the desired order in $\epsilon$.
The algebraic relations between masters with numerators and dotted master integrals sometimes contain 
coefficients with additional singularities in $\epsilon$.
This is a well-known feature in the algebraic reductions based on integrations by part
requiring a deeper expansion of the corresponding master integrals in $\epsilon$.
The complete set of master integrals needed, some of them being quite lengthy, contains HPLs up to weight six.
We refrain from printing the explicit results in this paper. 
They can be found at the webpage~\cite{Zeuthen-CAS:2009} in the file 
\href{http://www-zeuthen.desy.de/theory/research/supporting_files/Masters_V_QCD_2loops_eps.m}
{\ttfamily{Masters\_V\_QCD\_2loops\_eps.m}}.

Let us briefly describe the checks on the master integrals.
As was mentioned, we used differential equations to compute 
the analytic $x$-dependence of all coefficients of the Laurent expansions  
in $\epsilon$ in terms of HPLs. 
Numerical evaluations of the HPLs may be done with the {\ttfamily Mathematica} library \texttt{HPL}~\cite{Maitre:2005uu}.
In addition, we evaluated the master integrals 
at selected kinematical points in the Euclidean region by two independent methods.
To that end we (i) used the sector decomposition package \texttt{sector\_decomposition}~\cite{Bogner:2007cr} 
and (ii) derived and evaluated  Mellin-Barnes representations 
for the masters with the programs \texttt{ambre}~\cite{Gluza:2007rt} and \texttt{MB}~\cite{Czakon:2005rk}.

In principle, the method of differential equations allows the computation of master integrals to any depth in $\ep$.  
In practice, though, every additional power of $\ep$ poses a new challenge. 
Nevertheless, we have been able to push the calculation for each of the master integrals 
even to one power in $\ep$ beyond what is actually needed (see Tab.\ref{tab:MIs}).
For instance, the on-shell master integral {\ttfamily SE3l3mOS} (see Fig.~\ref{fig:MIsSE}) 
has been evaluated until terms of the order ${\cal O}(\epsilon^4)$. 
Due to the iterative nature of the method of differential equations this implies
further constraints and an additional self-consistency check on the
correctness of the solutions.
As a consequence of our strategy we met at intermediate steps HPLs up to weight seven.
We observe the cancellation of all HPLs with weights bigger than five at order ${\cal O}(\epsilon)$ 
after summation of all two-loop Feynman diagrams in Fig.~\ref{fig:massiveFF2}.
An analogous statement applies to the corresponding constants of weight six. 
All terms like e.g. $\zeta_6$, $\zeta_3^2$, or $s_6 = S(-5,-1; \infty) = 0.9874414264032997137...$ 
(the latter one is defined e.g. in Ref.~\cite{Maitre:2005uu}) disappear in the final result.
The master integrals and their Laurent series are now known sufficiently deep
in $\ep$ for a computation of the QCD form factor at two loops including all terms of order $\ep^2$. 
This is clearly beyond the scope of the present paper, but we comment on this
issue again in Sec.~\ref{sec:appl}.

\subsection{Renormalization}
From the Feynman diagrams of Figs.~\ref{fig:massiveFF1} and~\ref{fig:massiveFF2} 
we obtain the bare results for ${\cal F}_1^{\rm{b}}$ and ${\cal F}_2^{\rm{b}}$
within dimensional regularization, $D=4-2\ep$. 
The ultraviolet divergences require renormalization by adding 
the appropriate counter-terms $CT_1$ and $CT_2$ and 
we briefly present all necessary formulae.
The relevant parameters to be renormalized are the strong coupling constant $\alpha_{s}$, the
external (heavy-quark) wave function $\psi$ and the heavy-quark mass $m$. 
The latter denotes the pole mass, so that the renormalization of $m$ imposes the on-shell condition.
The bare (unrenormalized) quantities are multiplicatively related to the renormalized ones, 
\begin{eqnarray}
\label{eq:Z-renorm} 
\alpha_{s}^{\rm{b}} \, = \, Z_{\alpha_{s}}\, \alpha_{s}\, ,
\qquad\qquad
m^{\rm{b}} \, = \, Z_{m}\, m\, ,
\qquad\qquad
\psi^{\rm{b}} & = & \sqrt{Z_{2}}\, \psi\, ,
\end{eqnarray}
and we work in the $\MSbar$-scheme, but $Z_{2}$ and $Z_{m}$ are to be taken in the on-shell scheme.
The necessary renormalization constants are all known~\cite{Bernreuther:2004ih,Melnikov:2000qh,Melnikov:2000zc} 
and through two loops also to sufficient depth in $\epsilon$,
{\small 
\begin{eqnarray}
\label{eq:alphas-Z1l}
  Z_{\alpha_s}^{(1)} &=& - {\beta_0 \over \epsilon} 
  \, ,
\\
\label{eq:mass-Z1l}
  Z_{m}^{(1)} &=& 
  \Gamma(1+\ep) \* \left({\mu^2 \over m^2}\right)^\ep \*
  \biggl\{
  - \cf \* \left( {3 \over \ep} + {4 \over 1 - 2 \* \ep} \right)
  \biggr\}
  \, ,
\\
\label{eq:wave-Z21l}
  Z_{2}^{(1)} &=& 
  \Gamma(1+\ep) \* \left({\mu^2 \over m^2}\right)^\ep \*
  \biggl\{
  - \cf \* \left( {3 \over \ep} + {4 \over 1 - 2 \* \ep} \right)
  \biggr\}
  \, ,
\\
\label{eq:wave-Z22l}
  Z_{2}^{(2)} &=& 
  \Gamma(1+\ep)^2 \* \left({\mu^2 \over m^2}\right)^{2\*\ep} \*
  \biggl\{
         {1 \over \ep^2} \* \cf \* \biggl( 
         {9 \over 2} \* \cf
       + {11 \over 2} \* \ca 
       - 2 \* \tr \* \nl
          \biggr)
       + {1 \over \ep} \* \cf \* \biggl( 
         {51 \over 4} \* \cf 
       - {127 \over 12} \* \ca
       + {11 \over 3} \* \tr \* \nl
       + \tr
          \biggr)
\\
& &\mbox{}
       + \cf^2  \*  \biggl(
            {433 \over 8}
          - 24 \* \z3
          - 78 \* \z2
          + 96 \* \z2 \* \lntwo
          \biggr)
       + \cf \* \ca  \*  \biggl(
          - {1705 \over 24}
          + 12 \* \z3
          + {49 \over 2} \* \z2
          - 48 \* \z2 \* \lntwo
          \biggr)
\nonumber\\
& &\mbox{}
       + \cf \* \tr \* \nl  \*  \biggl(
            {113 \over 6}
          + 10 \* \z2
          \biggr)
       + \cf \* \tr  \*  \biggl(
            {947 \over 18}
          - 30 \* \z2
          \biggr)
\nonumber\\
& &\mbox{}
       + \ep \* \cf \* \biggl[
         \cf  \*  \biggl(
            {211 \over 16}
          - 267 \* \z2
          - 294 \* \z3
          + 552 \* \z2 \* \lntwo
          - 384 \* \a4
          - 16 \* \lntwof
          - 192 \* \z2 \* \lntwos
          + {1008 \over 5} \* \z2^2
          \biggr)
\nonumber\\
& &\mbox{}
       + \ca  \*  \biggl(
          - {9907 \over 48}
          + {224 \over 3} \* \z2
          + {398 \over 3} \* \z3
          - 276 \* \z2 \* \lntwo
          + 192 \* \a4
          + 8 \* \lntwof
          + 96 \* \z2 \* \lntwos
          - {504 \over 5} \* \z2^2
          \biggr)
\nonumber\\
& &\mbox{}
       + \tr \* \nl  \*  \biggl(
            {851 \over 12}
          + {116 \over 3} \* \z2
          + {44 \over 3} \* \z3
          \biggr)
       + \tr  \*  \biggl(
            {17971 \over 108}
          - {448 \over 3} \* \z2
          - {340 \over 3} \* \z3
          + 192 \* \z2 \* \lntwo
          \biggr)
          \biggr]
  \biggr\}
\nonumber
  \, ,
\end{eqnarray}
}
where we have always set the factor $S_\epsilon=(4 \pi)^\ep \exp(-\ep \gamma_{\rm E}) = 1$ and 
in the $\MSbar$-scheme $\beta_0 = 11/3 \ca - 2/3 \nf$.

In addition to Eqs.~(\ref{eq:alphas-Z1l})--(\ref{eq:wave-Z22l}) the two-loop counter-terms $CT_i^{(2)}$, $i=1,2$ 
contain the subtraction of those one-loop sub-divergences from the two-loop diagrams of Fig.~\ref{fig:massiveFF2} 
which are related to the on-shell renormalization of the heavy mass $m$ in the loop.
At one loop we denote the relevant results by ${\cal F}_1^{{\rm{mct}},(1)}$ and ${\cal F}_2^{{\rm{mct}},(1)}$.
They can be derived from diagram $One$ in Fig.~\ref{fig:massiveFF1} 
upon insertion of a one-loop mass counter-term for the heavy quark in the loop (see~\cite{Bernreuther:2004ih}).
${\cal F}_1^{{\rm{mct}},(1)}$ and ${\cal F}_2^{{\rm{mct}},(1)}$ need to be
calculated to order ${\cal O}(\ep^2)$ and we find in space-like kinematics,
$q^2 = - Q^2 < 0$, the following results 
{\small
\begin{eqnarray}
  \label{eq:F1mctl1}
{\cal F}_1^{{\rm{mct}},(1)} &=&
         {1 \over \ep} \* \cf  \*  \biggl\{
            (1 - 2 \* (\x3) + 3 \* \x2 - 3 \* \x1) \* (2 \* H_{0})
          + 2 \* H_{0} \* p(x)
          - 2 \* (1 + 2 \* \x2 - 2 \* \x1)
          \biggr\}
\\
&+&\mbox{}
         \cf  \*  \biggl\{
            4
          + ( - 2 \* \z2 - 4 \* H_{-1,0} + 2 \* H_{0} + 2 \* H_{0,0}) \* p(x)
          + ( - 2 \* \z2 - 4 \* H_{-1,0} + 2 \* H_{0,0}) \* (1 - 2 \* (\x3) + 3 \* \x2 - 3 \* \x1)
\nonumber\\
& &\mbox{}
          + 2 \* H_{0} \* (1 - 4 \* (\x3) + 6 \* \x2 - 4 \* \x1)
          \biggr\}
\nonumber\\
&+&\mbox{}
         \ep \* \cf  \*  \biggl\{
            4 \* H_{0} \* (1 - 6 \* (\x3) + 9 \* \x2 - 5 \* \x1)
          - 8 \* (1 + 3 \* \x2 - 3 \* \x1)
            + ( - 4 \* \z3 - 2 \* \z2 + 4 \* H_{-1} \* \z2 + 8 \* H_{-1,-1,0}
\nonumber\\
& &\mbox{}
            - 4 \* H_{-1,0} 
            - 4 \* H_{-1,0,0} + 4 \* H_{0} - H_{0} \* \z2 - 4 \* H_{0,-1,0}
	    + 2 \* H_{0,0} 
            + 2 \* H_{0,0,0}) \* p(x)
          + ( - 4 \* \z3 + 4 \* H_{-1} \* \z2 
\nonumber\\
& &\mbox{}
            + 8 \* H_{-1,-1,0} 
            - 4 \* H_{-1,0,0}
            - H_{0} \* \z2 - 4 \* H_{0,-1,0} + 2 \* H_{0,0,0}) \* (1 - 2 \* (\x3) + 3 \* \x2 - 3 \* \x1)
          - 3 \* \z2 \* (1 - 8/3 \* (\x3) 
\nonumber\\
& &\mbox{}
          + 14/3 \* \x2 
          - 10/3 \* \x1)
          - ( 4 \* H_{-1,0} - 2 \* H_{0,0}) \* (1 - 4 \* (\x3) + 6 \* \x2 - 4 \* \x1)
          \biggr\}
\nonumber\\
&+&\mbox{}
         \ep^2 \* \cf  \*  \biggl\{
            16 \* (1 + \x2 - \x1)
          + ( - 4 \* \z3 - 4 \* \z2 - 14/5 \* \z2^2 + 8 \* H_{-1} \* \z3 + 4 \* H_{-1} \* \z2 - 8 \* H_{-1,-1} \* \z2 - 16 \* H_{-1,-1,-1,0} 
\nonumber\\
& &\mbox{}
	    + 8 \* H_{-1,-1,0} + 8 \* H_{-1,-1,0,0} - 8 \* H_{-1,0} + 2 \* H_{-1,0} \* \z2 + 8 \* H_{-1,0,-1,0} 
	    - 4 \* H_{-1,0,0} - 4 \* H_{-1,0,0,0} + 8 \* H_{0} 
\nonumber\\
& &\mbox{}
            - 14/3 \* H_{0} \* \z3 
	    - H_{0} \* \z2 + 4 \* H_{0,-1} \* \z2 + 8 \* H_{0,-1,-1,0} - 4 \* H_{0,-1,0} 
	    - 4 \* H_{0,-1,0,0} + 4 \* H_{0,0} - H_{0,0} \* \z2 - 4 \* H_{0,0,-1,0} 
\nonumber\\
& &\mbox{}
	    + 2 \* H_{0,0,0} + 2 \* H_{0,0,0,0}) \* p(x)
          - 2 \* \z2 \* (1 - 12 \* (\x3) + 18 \* \x2 - 10 \* \x1)
          - 10/3 \* \z3 \* (1 - 24/5 \* (\x3) + 34/5 \* \x2 
\nonumber\\
& &\mbox{}
          - 22/5 \* \x1)
          + ( - 14/5 \* \z2^2 + 8 \* H_{-1} \* \z3 - 8 \* H_{-1,-1} \* \z2 - 16 \* H_{-1,-1,-1,0} 
	    + 8 \* H_{-1,-1,0,0} + 2 \* H_{-1,0} \* \z2 
\nonumber\\
& &\mbox{}
            + 8 \* H_{-1,0,-1,0} - 4 \* H_{-1,0,0,0}
	    - 14/3 \* H_{0} \* \z3 + 4 \* H_{0,-1} \* \z2 + 8 \* H_{0,-1,-1,0} - 4 \* H_{0,-1,0,0}
	    - H_{0,0} \* \z2 
\nonumber\\
& &\mbox{}
            - 4 \* H_{0,0,-1,0} + 2 \* H_{0,0,0,0}) \* (1 - 2 \* (\x3) + 3 \* \x2 - 3 \* \x1)
          + (4 \* H_{-1} \* \z2 + 8 \* H_{-1,-1,0} - 4 \* H_{-1,0,0} + 8 \* H_{0} 
	    - H_{0} \* \z2 
\nonumber\\
& &\mbox{}
            - 4 \* H_{0,-1,0} + 2 \* H_{0,0,0}) \* (1 - 4 \* (\x3) + 6 \* \x2 - 4 \* \x1)
          - ( 8 \* H_{-1,0} - 4 \* H_{0,0}) \* (1 - 6 \* (\x3) + 9 \* \x2 - 5 \* \x1)
          \biggr\}
\nonumber
       + \calO(\ep^3)
\, ,
\\
  \label{eq:F2mctl1}
{\cal F}_2^{{\rm{mct}},(1)} &=& 
         {1 \over \ep} \* \cf  \*  \biggl\{
            4 \* (\x2 - \x1)
          + 4 \* H_{0} \* (\x3 - 3/2 \* \x2 + 1/2 \* \x1)
          - H_{0} \* q(x)
          \biggr\}
\\
&+&\mbox{}
         \cf  \*  \biggl\{
            8 \* (\x2 - \x1)
          + ( 16 \* H_{0} + 4 \* H_{0,0} - 4 \* \z2 - 8 \* H_{-1,0}) \* (\x3 - 3/2 \* \x2 + 1/2 \* \x1)
          + (z2 + 2 \* H_{-1,0} - H_{0,0}) \* q(x)
          \biggr\}
\nonumber\\
&+&\mbox{}
         \ep \* \cf  \*  \biggl\{
            8 \* (\x2 - \x1)
          + ( 8 \* H_{-1} \* \z2  - 8 \* \z3 + 16 \* H_{-1,-1,0} - 32 \* H_{-1,0}
            - 8 \* H_{-1,0,0} + 24 \* H_{0} - 2 \* H_{0} \* \z2 - 8 \* H_{0,-1,0} 
\nonumber\\
& &\mbox{}
            + 16 \* H_{0,0} + 4 \* H_{0,0,0}) \* (\x3 - 3/2 \* \x2 + 1/2 \* \x1)
          + (2 \* \z3 - 2 \* H_{-1} \* \z2 - 4 \* H_{-1,-1,0} + 2 \* H_{-1,0,0} + 1/2 \* H_{0} \* \z2 
\nonumber\\
& &\mbox{}         
          + 2 \* H_{0,-1,0} - H_{0,0,0}) \* q(x)
          - 16 \* \z2 \* (\x3 - 13/8 \* \x2 + 5/8 \* \x1)
          \biggr\}
\nonumber\\
&+&\mbox{}
         \ep^2 \* \cf  \*  \biggl\{
            16 \* (\x2 - \x1)
          - 24 \* \z2 \* (\x3 - 5/3 \* \x2 + 2/3 \* \x1)
          - 32 \* \z3 \* (\x3 - 35/24 \* \x2 + 11/24 \* \x1)
\nonumber\\
& &\mbox{}         
          + ( - 28/5 \* \z2^2 + 16 \* H_{-1} \* \z3 + 32 \* H_{-1} \* \z2 - 16 \* H_{-1,-1} \* \z2 
	    - 32 \* H_{-1,-1,-1,0} + 64 \* H_{-1,-1,0} 
	    + 16 \* H_{-1,-1,0,0} 
\nonumber\\
& &\mbox{} 
            - 48 \* H_{-1,0} + 4 \* H_{-1,0} \* \z2 + 16 \* H_{-1,0,-1,0} 
	    - 32 \* H_{-1,0,0} - 8 \* H_{-1,0,0,0} + 32 \* H_{0} - 28/3 \* H_{0} \* \z3 
	    - 8 \* H_{0} \* \z2 
\nonumber\\
& &\mbox{}         
            + 8 \* H_{0,-1} \* \z2 + 16 \* H_{0,-1,-1,0} - 32 \* H_{0,-1,0} 
	    - 8 \* H_{0,-1,0,0} + 24 \* H_{0,0} - 2 \* H_{0,0} \* \z2 - 8 \* H_{0,0,-1,0} 
	    + 16 \* H_{0,0,0} 
\nonumber\\
& &\mbox{}         
            + 4 \* H_{0,0,0,0}) \* (\x3 - 3/2 \* \x2 + 1/2 \* \x1)
          + (7/5 \* \z2^2 - 4 \* H_{-1} \* \z3 + 4 \* H_{-1,-1} \* \z2 + 8 \* H_{-1,-1,-1,0} 
	    - 4 \* H_{-1,-1,0,0} 
\nonumber\\
& &\mbox{}         
            - H_{-1,0} \* \z2 - 4 \* H_{-1,0,-1,0}
            + 2 \* H_{-1,0,0,0} - 4 \* H_{0} + 7/3 \* H_{0} \* \z3 - 2 \* H_{0,-1} \* \z2
            - 4 \* H_{0,-1,-1,0} + 2 \* H_{0,-1,0,0} 
\nonumber\\
& &\mbox{}         
            + 1/2 \* H_{0,0} \* \z2 
	    + 2 \* H_{0,0,-1,0} - H_{0,0,0,0}) \* q(x)
          \biggr\}
\nonumber
       + \calO(\ep^3)
\, .
\qquad
\end{eqnarray}
}
${\cal F}_1^{{\rm{mct}},(1)}$ and ${\cal F}_2^{{\rm{mct}},(1)}$ have been computed before to order ${\cal O}(\ep)$ (see~\cite{Bernreuther:2004ih}),  
the ${\cal O}(\ep^2)$ terms are new.\footnote{%
  There are some misprints in Eqs.~(37) and (38) of Ref.~\cite{Bernreuther:2004ih}. 
  However the source files of the authors agree with the corresponding parts of our Eqs.~(\ref{eq:F1mctl1}) and~(\ref{eq:F2mctl1}).}
All results are given in terms of HPLs with argument $0 \le x \le 1$.
Moreover, it is convenient to abbreviate the ubiquitous polynomial 
\begin{eqnarray}
\label{eq:pofx-def}
p(x) 
&\equiv& {1 + x^2 \over 1 - x^2} 
\, = \, 
{1 \over 2} \left(p_{\rm{qq}}(x) + p_{\rm{qq}}(-x) \right)
,
\end{eqnarray}
which is related to the quark-quark splitting function at one-loop,
$p_{\rm qq}(x) = 2/(1-x) - 1 - x$.
We use also the additional definitions
\begin{eqnarray}
\label{eq:xn-def}
x_n &=& {1 \over (1+x)^n}
\, , 
\\
\label{eq:qofx-def}
q(x)&=&  {1 \over 1-x}-{1 \over 1+x} 
\, .
\end{eqnarray}

With these ingredients we arrive at the following results for counter-terms at
one- and two-loops, $CT_i^{(1)}$ and $CT_i^{(2)}$.
These counter-terms need to be added to the sum of the unsubtracted (bare) diagrams 
in order to arrive at the (ultraviolet) renormalized heavy-quark form
factors.
\begin{eqnarray}
\label{eq:ct1loop}
  CT^{(1)}_i &=& Z_{2}^{(1)} \, {\cal F}_i^{(0)} \, 
  \, ,
\\
\label{eq:ct2loop}
  CT^{(2)}_i &=& 
  - 2\, Z_{m}^{(1)} \, {\cal F}_i^{{\rm{mct}},(1)} \, 
  + \left( Z_{\alpha_s}^{(1)} + Z_{2}^{(1)} 
    \right) \, {\cal F}_i^{{\rm{b}},(1)}\, 
  + Z_{2}^{(2)} \, {\cal F}_i^{(0)}
  \, ,
\end{eqnarray}
where $i=1,2$ and, 
trivially, ${\cal F}_1^{(0)} = 1$ and ${\cal F}_2^{(0)} = 0$ at the Born level.
The bare results at one loop ${\cal F}_i^{{\rm{b}},(1)}$ are needed to order
${\cal O}(\ep^2)$ as well. We do not display them here, 
but the interested reader can easily derive them with the help of 
Eqs.~(\ref{eq:ct1loop}) and (\ref{eq:Frendef}) from the renormalized result 
${\cal F}_i^{(1)}$ presented in the next Section.
To order ${\cal O}(\ep)$ the explicit expressions 
for ${\cal F}_1^{{\rm{b}},(1)}$ and ${\cal F}_2^{{\rm{b}},(1)}$ are also given in~\cite{Bernreuther:2004ih}.

%
%
%
\section{Results}
\label{sec:results}
%
%
We are now in a position to present our main results. 
In space-like kinematics, $q^2 = - Q^2 < 0$, 
the renormalized form factors ${\cal F}_1$ (electric) 
and ${\cal F}_2$ (magnetic) through NNLO in QCD and up to ${\cal O}(\epsilon)$
are expressed in terms of HPLs of argument $x$, $0 \le x \le 1$, 
see Eq.~(\ref{eq:spacelike-xdef}).
We also factorize again the polynomials $p(x)$, $q(x)$ and employ the
abbreviations $x_n$, see Eqs.~(\ref{eq:pofx-def})--(\ref{eq:qofx-def}).
Adding the bare diagrams in Figs.~\ref{fig:massiveFF1} and \ref{fig:massiveFF2}
together with the counter-terms of Eqs.~(\ref{eq:ct1loop}) and (\ref{eq:ct2loop}),
\begin{eqnarray}
\label{eq:Frendef}
{\cal F}_i^{(l)} &=& {\cal F}_i^{{\rm{b}},(l)} + CT^{(l)}_i
\, ,
\end{eqnarray}
and setting the scale to $\mu^2 = m^2$, we have:
{\small
\begin{eqnarray}
  \label{eq:F1rl1}
{\lefteqn{
{\cal F}_1^{(1)} \,=\,}}
\\
&&
       {1 \over \ep} \* \cf \* \biggl\{
          - 2
          - 2 \* p(x) \* H_{0}
        \biggr\}
\nonumber\\
       &-&\mbox{} \cf \* \biggl\{
            4
          + (1 - 2 \* x_1) \* H_{0}
          + 2 \* p(x) \* (
            H_{0,0}
          - 2 \* H_{-1,0}
          - \z2
          + 2 \* H_{0}
          )
        \biggr\}
\nonumber\\
&-&\mbox{}
         \ep \* \cf \* \biggl\{
            8
          + p(x) \* (
            8 \* H_{0}
          + 8 \* H_{-1,-1,0}
          - 4 \* \z3
          + 4 \* H_{-1} \* \z2
          - 4 \* H_{-1,0,0}
          - H_{0} \* \z2
          - 4 \* H_{0,-1,0}
          + 2 \* H_{0,0,0}
          + 4 \* H_{0,0}
\nonumber\\
& &\mbox{}
          - 8 \* H_{-1,0}
          - 4 \* \z2
          )
          + (1 - 2 \* x_1 ) \* ( 
            H_{0,0}
          - 2 \* H_{-1,0}
          )
          + 2 \* x_1 \* \z2 
        \biggr\}
\nonumber\\
&-&\mbox{}
         \ep^2 \* \cf \* \biggl\{
            16
          + 2 \* \z2
          - 2/3 \* \z3
          + p(x) \* (
          - 8 \* \z2
          - 8 \* \z3
          - 7 \* \z4
          + 8 \* H_{-1} \* \z3
          - 8 \* H_{-1,-1} \* \z2
          - 16 \* H_{-1,-1,-1,0}
          + 8 \* H_{-1,-1,0,0}
\nonumber\\
& &\mbox{}
          + 8 \* H_{-1,0,-1,0}
          + 2 \* H_{-1,0} \* \z2
          - 4 \* H_{-1,0,0,0}
          + 16 \* H_{0}
          + 8 \* H_{-1} \* \z2
          + 16 \* H_{-1,-1,0}
          - 8 \* H_{-1,0,0}
          - 2 \* H_{0} \* \z2
          - 8 \* H_{0,-1,0}
\nonumber\\
& &\mbox{}
          + 4 \* H_{0,0,0}
          - 14/3 \* H_{0} \* \z3
          + 4 \* H_{0,-1} \* \z2
          + 8 \* H_{0,-1,-1,0}
          - 4 \* H_{0,-1,0,0}
          - 16 \* H_{-1,0}
          - H_{0,0} \* \z2
          - 4 \* H_{0,0,-1,0}
          + 2 \* H_{0,0,0,0}
\nonumber\\
& &\mbox{}
          + 8 \* H_{0,0}
          )
             + ( 1 - 2 \* x_1 ) \* (
            2 \* H_{-1} \* \z2
          + 4 \* H_{-1,-1,0}
          - 2 \* H_{-1,0,0}
          - 1/2 \* H_{0} \* \z2
          - 2 \* \z3
          - 2 \* H_{0,-1,0}
          + H_{0,0,0}
          )
        \biggr\}
       + \calO(\ep^3)
\, ,\nonumber
\end{eqnarray}
%
%
\oddsidemargin -0.6cm 

%
%
%
\begin{eqnarray}
{\lefteqn{
\label{eq:F2rl1}
{\cal F}_2^{(1)} \,=\,
}}
\\
&&
      \cf \* \biggl\{ - 2\*q(x) \* H_{0} \biggl\}
\nonumber\\
&+&\mbox{} 
        \ep\* \cf \* \biggl\{ q(x) 
         \* (2\* \z2 + 4\*H_{-1,0} - 8\*H_{0} - 2\*H_{0,0})
          \biggr\}
\nonumber\\
&+&\mbox{} 
        \ep^2\*\cf \* \biggr\{
          q(x)\* (4\*\z3 + 8\*\z2 - 4\*H_{-1}\*\z2 - 8\*H_{-1,-1,0} + 
16\*H_{-1,
         0} + 4\*H_{-1,0,0} - 16\*H_{0} + H_{0}\*\z2 + 4\*H_{0,-1,0}
\nonumber\\
& &\mbox{}
          - 8\*H_{0,0} - 2\*H_{0,0,0})
          \biggr\}
       + \calO(\ep^3)\nonumber
\, ,
\end{eqnarray}
%
%
%

}
\oddsidemargin -0.1cm 

The one-loop part of ${\cal F}_1$ in Eq.~(\ref{eq:F1rl1}) 
has first been given in Ref.~\cite{Mitov:2006xs}, 
and also ${\cal F}_2^{(1)}$ in Eq.~(\ref{eq:F2rl1}) was known to $\calO (\ep)$ before~\cite{Bernreuther:2004ih}.
For the two-loop parts of Eqs.~(\ref{eq:F1rl2}) and~(\ref{eq:F2rl2}) up to 
$\calO (\epsilon^0)$ we have found agreement with Ref.~\cite{Bernreuther:2004ih}.
All other terms, especially those at $\calO (\epsilon)$ in Eqs.~(\ref{eq:F1rl2}) and (\ref{eq:F2rl2}), are new.

In time-like kinematics above production threshold ($Q^2 > 4 m^2$) the conformal variable $y$ is defined as 
\begin{equation}
  \label{eq:timelike-xdef}
  \displaystyle
  y = {\sqrt{Q^2} - \sqrt{Q^2 - 4m^2} \over \sqrt{Q^2} + \sqrt{Q^2 - 4m^2}} \, .
\end{equation}
From Eqs.~(\ref{eq:F1rl1})--(\ref{eq:F2rl2}) the corresponding results 
for the form factor can be obtained by means of a suitable analytic continuation, 
$x \to y = -x$, taking into account the (complex) continuation of $Q^2$.
This continuation is easily performed with the help of 
routines for HPLs \cite{Remiddi:1999ew} 
implemented in {\ttfamily FORM}~\cite{Vermaseren:2000nd}, 
the only subtle point being logarithmic branch cuts, starting with 
\begin{equation}
  \label{eq:lnxcnt}
  \ln x \;\rightarrow\; \ln y + {\rm i}\,\pi \:\: .
\end{equation}
Thanks to Eq.~(\ref{eq:lnxcnt}), both ${\cal F}_1$ and ${\cal F}_2$ develop an
imaginary part in the time-like region, 
\begin{equation}
  \label{eq:tmlkdef}
  {\cal F}_i \,=\,
  \Re {\cal F}_i 
  + {\rm i}\, \Im {\cal F}_i 
  \, .
\end{equation}

\subsection{Asymptotic expansions}
Let us next study the asymptotic expansions of Eqs.~(\ref{eq:F1rl1})--(\ref{eq:F2rl2}).
In the (space-like) high-energy limit, $Q^2 \gg m^2$, i.e. $x \to 0$, 
we abbreviate the logarithmically
enhanced terms, 
\begin{eqnarray}
  \label{eq:L-def}
  L = \ln\left( {Q^2\over m^2}\right)\, .
\end{eqnarray}
Then, employing algebraic properties of HPLs 
and keeping the scale $\mu^2 = m^2$ 
we obtain from Eqs.~(\ref{eq:F1rl1})--(\ref{eq:F2rl2}) the asymptotic expansions for the renormalized form factors. 
Here and in the following we set $\tr = 1/2$.
{\small
\begin{eqnarray}
  \label{eq:F1rl1-x0}
{\cal F}_1^{(1)} &=&
         \cf \* \biggl\{
         {1 \over \ep} \* (
            2 \* L
          - 2
         )
          - L^2
          + 3 \* L
          - 4
          + 2 \* \z2
       + \ep \* \biggl(
            {1 \over 3} \* L^3
          - {3 \over 2} \* L^2
          + (
            8
          - \z2 ) \* L
          - 8
          + 2 \* \z2
          + 4 \* \z3
       \biggr)
\\
& &\mbox{}
       + \ep^2 \* \biggl(
          - {1 \over 12} \* L^4
          + {1 \over 2} \* L^3
          - \biggl(
            4
          - {1 \over 2} \* \z2 \biggr) \* L^2
          + \biggl(
            16
          - {3 \over 2} \* \z2
          - {14 \over 3} \* \z3 \biggr) \* L
          - 16
          + 6 \* \z2
          + {20 \over 3} \* \z3
          + {14 \over 5} \* \z2^2
       \biggr)
       \biggr\}
       + \calO(\ep^3)
\nonumber\, ,
\\[1ex]
\label{eq:F1rl2-x0}
{\cal F}_1^{(2)} &=&
         \cf^2 \* \biggl\{
         {1 \over \ep^2} \* (
            2 \* L^2
          - 4 \* L
          + 2
          )
       + {1 \over \ep} \* (
          - 2 \* L^3
          + 8 \* L^2
          - (
            14
          - 4 \* \z2 ) \* L
          + 8
          - 4 \* \z2
          )
          + {7 \over 6} \* L^4
          - {20 \over 3} \* L^3
\\
& &\mbox{}
          + \biggl(
            {55 \over 2}
          - 4 \* \z2
            \biggr) \* L^2
          - \biggl(
            {85 \over 2}
          - 32 \* \z3
            \biggr) \* L
          + 46
          + 39 \* \z2
          - 44 \* \z3
          - 48 \* \z2 \* \lntwo
          - {118 \over 5} \* \z2^2
       + \ep \* \biggl(
          - {1 \over 2} \* L^5
          + {11 \over 3} \* L^4
\nonumber\\
& &\mbox{}
          - \biggl(
            {137 \over 6}
          - {8 \over 3} \* \z2
            \biggr) \* L^3
          + \biggl(
            {153 \over 2}
          - {112 \over 3} \* \z3
            \biggr) \* L^2
          - \biggl(
            {479 \over 4}
          + 17 \* \z2
          - {284 \over 3} \* \z3
          - {106 \over 5} \* \z2^2
            \biggr) \*  L
          + 4
          + 163 \* \z2
          - {346 \over 3} \* \z3
\nonumber\\
& &\mbox{}
          - 24 \* \z2 \* \lntwo
          + 96 \* \z2 \* \lntwos
          - 160 \* \z2^2
          + 8 \* \lntwof
          + 192 \* \a4
          - 12 \* \z2 \* \z3
          - 18 \* \z5
          \biggr)
       \biggr\}
\nonumber\\
&+&\mbox{}
         \ca \* \cf \* \biggl\{
         {1 \over \ep^2} \* \biggl(
          - {11 \over 3} \* L
          + {11 \over 3}
          \biggr)
       + {1 \over \ep} \* \biggl(
          \biggl(
            {67 \over 9}
          - 2 \* \z2
          \biggr) \* L
          - {49 \over 9}
          + 2 \* \z2
          - 2 \* \z3
          \biggr)
          + {11 \over 9} \* L^3
          - \biggl(
            {233 \over 18}
          - 2 \* \z2
            \biggr) \* L^2
\nonumber\\
& &\mbox{}
          + \biggl(
            {2545 \over 54}
          + {22 \over 3} \* \z2
          - 26 \* \z3
            \biggr) \* L
          - {1595 \over 27}
          - {7 \over 9} \* \z2
          + {134 \over 3} \* \z3
          + 24 \* \z2 \* \lntwo
          - {3 \over 5} \* \z2^2
       + \ep \* \biggl(
          - {11 \over 12} \* L^4
\nonumber\\
& &\mbox{}
          + \biggl(
            {565 \over 54}
          - {4 \over 3} \* \z2
            \biggr) \* L^3
          - \biggl(
            {3337 \over 54}
          + {11 \over 2} \* \z2
          - 26 \* \z3
            \biggr) \* L^2
          + \biggl(
            {70165 \over 324}
          + {575 \over 18} \* \z2
          - {260 \over 3} \* \z3
          - {88 \over 5} \* \z2^2
            \biggr) \* L
          - {28745 \over 162}
\nonumber\\
& &\mbox{}
          - {71 \over 27} \* \z2
          + {1577 \over 9} \* \z3
          + 12 \* \z2 \* \lntwo
          - 4 \* \lntwof
          - 96 \* \a4
          - 48 \* \z2 \* \lntwos
          + {637 \over 5} \* \z2^2
          - 2 \* \z2 \* \z3
          - 157 \* \z5
          \biggr)
       \biggr\}
\nonumber\\
&+&\mbox{}
         \cf \* \nl \* \biggl\{
         {1 \over \ep^2} \* \biggl(
            {2 \over 3} \* L
          - {2 \over 3}
          \biggr)
       + {1 \over \ep} \* \biggl(
          - {10 \over 9} \* L
          + {10 \over 9}
          \biggr)
          - {2 \over 9} \* L^3
          + {19 \over 9} \* L^2
          - \biggl(
            {209 \over 27}
          + {4 \over 3} \* \z2
            \biggr) \* L
          + {212 \over 27}
          - {14 \over 9} \* \z2
          - {8 \over 3} \* \z3
\nonumber\\
& &\mbox{}
       + \ep \* \biggl(
            {1 \over 6} \* L^4
          - {47 \over 27} \* L^3
          + \biggl(
            {281 \over 27}
          + \z2
            \biggr) \* L^2
          - \biggl(
            {5813 \over 162}
          + {37 \over 9} \* \z2
          - {8 \over 3} \* \z3
            \biggr) \* L
          + {2602 \over 81}
          - {88 \over 27} \* \z2
          - {140 \over 9} \* \z3
          - {48 \over 5} \* \z2^2
          \biggr)
      \biggr\}
\nonumber\\
&+&\mbox{}
         \cf \* \biggl\{
          - {2 \over 9} \* L^3
          + {19 \over 9} \* L^2
          - \biggl(
            {265 \over 27} 
          + 2 \* \z2 
            \biggr) \* L
          + {766 \over 27}
          - {4 \over 3} \* \z2
       + \ep \*  \biggl(
            {1 \over 6} \* L^4
          - {47 \over 27} \* L^3
          + \biggl(
            {281 \over 27}
          + \z2
            \biggr) \* L^2
\nonumber\\
& &\mbox{}
          - \biggl(
            {191 \over 6} 
          + 3 \* \z2 
          - {28 \over 9} \* \z3 
            \biggr) \* L
          + {2069 \over 27}
          - {808 \over 27} \* \z2
          + {112 \over 3} \* \z2 \* \lntwo
          - {92 \over 3} \* \z3
          - {4 \over 5} \* \z2^2
          \biggr)
      \biggr\}
       + \calO(\ep^2)
\nonumber\, ,
\\[1ex]
  \label{eq:F2rl1-x0}
{\cal F}_2^{(1)} &=&
         {m^2 \over Q^2} \* \cf \* \biggl\{
            4 \* L
       + \ep \* \biggl(
          - 2 \* L^2
          + 16 \* L
          + 4 \* \z2
          \biggr)
       + \ep^2 \* \biggl(
            {2 \over 3} \* L^3
          - 8 \* L^2
          + \biggl(
            32
          - 2 \* \z2
            \biggr) \* L
          + 16\* \z2
          + 8 \* \z3
          \biggr)
       \biggr\}
       + \calO(\ep^3)
\, ,
\\[1ex]
  \label{eq:F2rl2-x0}
{\cal F}_2^{(2)} &=&
         {m^2 \over Q^2} \* \cf^2 \* \biggl\{ 
         {1 \over \ep} \* \biggl(
            8 \* L^2
          - 8 \* L
          \biggr)
          - 8 \* L^3
          + 34 \* L^2
          - \biggl(
            62 
          - 48 \* \z2
            \biggr) \* L
          + 60 \* \z2
          - 192 \* \z2 \* \lntwo
          + 16 \* \z3
\\
& &\mbox{}
       + \ep \* \biggl(
            {14 \over 3} \* L^4
          - {94 \over 3} \* L^3
          + \biggl(
            74 
          - 16 \* \z2 
            \biggr) \* L^2
          - \biggl(
            249 
          - 116 \* \z2 
          - 192 \* \z3 
            \biggr) \* L
\nonumber\\
& &\mbox{}
          - 20
          + 492 \* \z2
          - 1152 \* \z2 \* \lntwo
          + 128 \* \z3
          + 32 \* \lntwof
          + 768 \* \a4
          + 384 \* \z2 \* \lntwos
          - {1192 \over 5} \* \z2^2
          \biggr)
       \biggr\}
\nonumber\\
&+&\mbox{}
         {m^2 \over Q^2} \* \ca \* \cf \* \biggl\{ 
            {2 \over 3} \* L^2
          + {346 \over 9} \* L
          + 12
          - {244 \over 3} \* \z2
          + 96 \* \z2 \* \lntwo
          + 80 \* \z3
       + \ep \* \biggl(
          - {2 \over 3} \* L^3
          - \biggl(
            {250 \over 9}
          + 8 \* \z2
            \biggr) \* L^2
\nonumber\\
& &\mbox{}
          + \biggl(
            {8057 \over 27}
          + 38 \* \z2
          - 48 \* \z3
            \biggr) \* L
          + 78
          - {1768 \over 9} \* \z2
          + 576 \* \z2 \* \lntwo
          - 264 \* \z3
          - 192 \* \z2 \* \lntwos
          + {1504 \over 5} \* \z2^2
          - 16 \* \lntwof
\nonumber\\
& &\mbox{}
          - 384 \* \a4
          \biggr)
       \biggr\}
\nonumber\\
&+&\mbox{}
         {m^2 \over Q^2} \* \cf \* \nl \* \biggl\{ 
            {4 \over 3} \* L^2
          - {100 \over 9} \* L
          - {8 \over 3} \* \z2
       + \ep \*  \biggl(
          - {4 \over 3} \* L^3
          + {148 \over 9} \* L^2
          - \biggl(
            {1922 \over 27} 
          + 4 \* \z2 
            \biggr) \* L
          - {296 \over 9} \* \z2
          - 16 \* \z3
          \biggr)
       \biggr\}
\nonumber\\
& &\mbox{}
       + {m^2 \over Q^2} \* \cf \* \biggl\{
            {4 \over 3} \* L^2
          - {100 \over 9} \* L
          + {136 \over 3}
          - 8 \* \z2
       + \ep \*  \biggl(
          - {4 \over 3} \* L^3
          + {148 \over 9} \* L^2
          - \biggl(
            {1922 \over 27} 
          + 4 \* \z2 
            \biggr) \* L
          + {764 \over 9}
          + {128 \over 9} \* \z2
\nonumber\\
& &\mbox{}
          + 64 \* \z2 \* \lntwo
          - {128 \over 3} \* \z3
          \biggr)
       \biggr\}
       + \calO(\ep^2)
\nonumber\, .
\end{eqnarray}
}
For the magnetic form factor ${\cal F}_2$ only the first power in $m^2/Q^2$ has been kept in the expansion. 
In Eqs.~(\ref{eq:F1rl2-x0}) and (\ref{eq:F2rl2-x0}), we have also separated 
the contribution  proportional to $\cf \tr$ coming from the heavy-quark loop in diagram $QL$. 
This part is finite after renormalization 
and proportional to a single power of $C_F$ only, thus it 
enters at ${\cal O}(\ep^0)$ and ${\cal O}(\ep^1)$ at two loops.
Moreover, due to the magnetic form factor ${\cal F}_2^{(1)}$ in Eq.~(\ref{eq:F2rl1-x0}) 
being finite also the terms proportional to the number of light quarks $\nl$ are finite in Eq.~(\ref{eq:F2rl2-x0}). 

The expansions~(\ref{eq:F1rl1-x0}) and (\ref{eq:F1rl2-x0}) for ${\cal F}_1$ 
in combination with the exponentiation of the heavy-quark form factor provide an independent
consistency check on the correctness of our result, in particular of the new terms of order
$\epsilon$ at two loops in Eq.~(\ref{eq:F1rl2}). 
A detailed discussion of this aspect will be presented in the next Section.

\bigskip

Let us instead now turn to the limit $Q^2 \sim 4 m^2$, i.e. $y \to 1$, 
which provides the threshold expansion in time-like kinematics, 
cf. Eq.~(\ref{eq:timelike-xdef}).
The relevant small parameter in which we expand here is the heavy-quark velocity $\beta$,
\begin{equation}
  \label{eq:beta-def}
\beta = \sqrt{1-\frac{4m^2}{Q^2}} \, . 
\end{equation}
In this particular limit, i.e. $\beta \sim 0$, the well-known Coulomb singularities appear.
The inverse powers of $\beta$ contributing to the form factor 
have already been directly addressed in investigations of non-relativistic QCD (e.g. \cite{Czarnecki:1997vz,Pineda:2006ri}).
Without repeating too much of the discussion in the literature, it is perhaps instructive, 
to study the anatomy of Coulomb two-loop corrections based on the individual diagrams which we have at our disposal.
To that end, we list below the individual results for the threshold expansion 
of the bare Feynman diagrams from Fig.~\ref{fig:massiveFF2} 
up to terms of ${\cal O}(\beta)$ and terms of ${\cal O}(\ep)$ 
as they enter in the computation of the bare form factor ${\cal F}_i^{{\rm b},(2)}$,
\begin{eqnarray}
  \label{eq:addFdiag}
  {\cal F}_i^{{\rm b},(2)} &\: =\: & 
  2 {{\cal F}_i^{(\mathrm S)}}
  + {{\cal F}_i^{(\mathrm QL)}}
  + {{\cal F}_i^{(\mathrm GL)}}
  + 2 {{\cal F}_i^{(\mathrm QV)}}
  + 2 {{\cal F}_i^{(\mathrm GV)}}
  + {{\cal F}_i^{(\mathrm C)}}
  + {{\cal F}_i^{(\mathrm L)}}
  \, ,
\end{eqnarray}
with $i=1,2$ and the symmetry factor of $2$ has been added for the diagrams $GV$, $QV$ and $S$.
It is also understood, that the respective color factor multiplying 
each diagram is taken from column ${\bf 1}$ in Tab.~\ref{tab:FFcolor}, 
i.e. for a color-singlet current: 
%
%
\begin{table}
  \centering
  \begin{tabular}{|c|c|c|c|c|}
    \hline
    diagram & ${\bf 1}$ & ${\bf 8}$ & result for ${\cal F}_1$ & result for ${\cal F}_2$\\
    \hline\hline
& & & & \\[-2ex]
    L  & $C_F^2$                              & $\left(C_F-{C_A \over 2}\right)^2$ 
& Eq.~(\ref{eq:diaL1}) & Eq.~(\ref{eq:diaL2}) \\[1ex]
    C  & $\left(C_F-{C_A \over 2}\right) C_F$ & $\left(C_F-{C_A \over 2}\right) \left(C_F-C_A\right)$ 
& Eq.~(\ref{eq:diaC1}) & Eq.~(\ref{eq:diaC2}) \\[1ex]
    QV & $\left(C_F-{C_A \over 2}\right) C_F$ & $\left(C_F-{C_A \over 2}\right)^2$ 
& Eq.~(\ref{eq:diaQV1}) & Eq.~(\ref{eq:diaQV2}) \\[1ex]
    S  & $C_F^2$                              & $\left(C_F-{C_A \over 2}\right) C_F$ 
& Eq.~(\ref{eq:diaS1}) & Eq.~(\ref{eq:diaS2}) \\[1ex]
    QL & $C_F$                                & $C_F-{C_A \over 2}$
& Eq.~(\ref{eq:diaQL1}) & Eq.~(\ref{eq:diaQL2}) \\[1ex]
    GL & $C_F C_A$                            & $\left(C_F-{C_A \over 2}\right) C_A$ 
& Eq.~(\ref{eq:diaGL1}) & Eq.~(\ref{eq:diaGL2}) \\[1ex]
    GV & ${1 \over 2} C_F C_A$                & ${1 \over 2} \left(C_F-{C_A \over 2}\right) C_A$ 
& Eq.~(\ref{eq:diaGV1}) & Eq.~(\ref{eq:diaGV2}) \\[2ex]
    \hline\hline
  \end{tabular}
  \caption{\small
    \label{tab:FFcolor}
    The color coefficients multiplying the result of the individual Feynman diagrams of
    the quark form factor (cf. Fig.~\ref{fig:massiveFF2}) for a current in the 
    singlet ${\bf 1}$ or octet ${\bf 8}$ representation 
    of the color-SU$(N_c)$ gauge group. 
    The scalar functions are given in the respective equations.
  }
\end{table}

{\small
\begin{eqnarray}
\label{eq:diaL1}
{{\cal F}_1^{(\mathrm L)}} &=& 
  {1\over \ep^2} \* \biggl\{
       - {3 \over 4} \* {1 \over \beta^2} \* \z2
       + {9 \over 8}
       - {3 \over 2} \* \z2
       - \I \* \pi \* {3 \over 4} \* {1 \over \beta}
       \biggr\}
  + {1\over \ep} \* \biggl\{
       {1 \over \beta^2} \* \biggl(
          - {3 \over 2} \* \z2
          + 3 \* \z2 \* \lntwo
          + 3 \* \z2 \* \lnbeta
          \biggr)
       + {3 \over 2} \* {1 \over \beta} \* \z2
       - {3 \over 16}
\\
& &\mbox{}
       - 3 \* \z2
       + 6 \* \z2 \* \lntwo
       + 6 \* \z2 \* \lnbeta 
       - \I \* \pi \* {3 \over 2} \* {1 \over \beta^2} \* \z2
       + \I \* \pi \* {1 \over \beta} \* \biggl(
            {3 \over 4}
          - {1 \over 2} \* \lntwo
          + {1 \over 2} \* \lnbeta
          \biggr)
       - 3 \* \I \* \pi \* \z2 
          \biggr\}
\nonumber\\
& &\mbox{}
       + {1 \over \beta^2} \* \biggl(
          - 7 \* \z2
          + 6 \* \z2 \* \lntwo
          - 6 \* \z2 \* \lntwos
          + {9 \over 2} \* \z2^2
          + 6 \* \z2 \* \lnbeta
          - 12 \* \z2 \* \lntwo \* \lnbeta
          - 6 \* \z2 \* \lnbetas
          \biggr)
\nonumber\\
& &\mbox{}
       + {1 \over \beta} \* \biggl(
          - {45 \over 2} \* \z2
          + 39 \* \z2 \* \lntwo
          + 21 \* \z2 \* \lnbeta
          \biggr)
          + {53 \over 96}
          - {179 \over 24} \* \z2
          - {63 \over 4} \* \z3
          - 15 \* \z2 \* \lntwo
          - 12 \* \z2 \* \lntwos
          + 9 \* \z2^2
\nonumber\\
& &\mbox{}
          - 6 \* \z2 \* \lnbeta
          - 24 \* \z2 \* \lntwo \* \lnbeta
          - 12 \* \z2 \* \lnbetas
       - \I \* \pi \* {1 \over \beta^2} \* \biggl(
            3 \* \z2
          - 6 \* \z2 \* \lntwo
          - 6 \* \z2 \* \lnbeta
          \biggr)
\nonumber\\
& &\mbox{}
       - \I \* \pi \* {1 \over \beta} \* \biggl(
            {3 \over 2}
          + {23 \over 2} \* \lntwo
          - {21 \over 2} \* \lntwos
          + {7 \over 2} \* \z2
          + {15 \over 2} \* \lnbeta
          - 13 \* \lntwo \* \lnbeta
          - {7 \over 2} \* \lnbetas
          \biggr)
\nonumber\\
& &\mbox{}
       + \I \* \pi  \*  (
            3 \* \z2
          + 12 \* \z2 \* \lntwo
          + 12 \* \z2 \* \lnbeta
          )
\, ,
\nonumber\\
\label{eq:diaL2}
{{\cal F}_2^{(\mathrm L)}} &=& 
    {1\over \ep} \* \biggl\{
       {3 \over 2} \* {1 \over \beta^2} \* \z2
       - {3 \over 2}
       + \I \* \pi \* {5 \over 4} \* {1 \over \beta}
          \biggr\}
       + {1 \over \beta^2} \* (
            7 \* \z2
          - 6 \* \z2 \* \lntwo
          - 6 \* \z2 \* \lnbeta
          )
       - {3 \over 2} \* {1 \over \beta} \* \z2
       - {37 \over 12}
       + {25 \over 12} \* \z2
       - {7 \over 4} \* \z3
       - 3 \* \z2 \* \lntwo
\\
& &\mbox{}
       - 2 \* \z2 \* \lnbeta
       + \I \* \pi \* 3 \* {1 \over \beta^2} \* \z2
       + \I \* \pi \* {1 \over \beta} \* \biggl(
            {3 \over 2}
          + {3 \over 2} \* \lntwo
          - {1 \over 2} \* \lnbeta
          \biggr)
       + \I \* \pi  \* \z2        
\, ,
\nonumber\\
\label{eq:diaC1}
{{\cal F}_1^{(\mathrm C)}} &=& 
  {1\over \ep} \* \biggl\{
         3 \* \z2 \* {1 \over \beta}
          - {3 \over 2}
       - \I \* \pi \* {1 \over \beta} \* (
            1
          - 2 \* \lntwo
          - \lnbeta
          )
          \biggr\}
       + {1 \over \beta} \* (
            24 \* \z2
          - 48 \* \z2 \* \lntwo
          - 30 \* \z2 \* \lnbeta
          )
       + {571 \over 60}
       + {269 \over 600} \* \z2
\\
& &\mbox{}
       + {137 \over 10} \* \z3
       + {122 \over 5} \* \z2 \* \lntwo
       + {44 \over 5} \* \z2 \* \lnbeta
       - \I \* \pi  \* {1 \over \beta} \* (
            5
          - 12 \* \lntwo
          + 12 \* \lntwos
          - 5 \* \z2
          - 8 \* \lnbeta
          + 16 \* \lntwo \* \lnbeta
          + 5 \* \lnbetas
          )
\nonumber\\
& &\mbox{}
       - \I \* \pi  \* {22 \over 5} \* \z2
\, ,
\nonumber\\
\label{eq:diaC2}
{{\cal F}_2^{(\mathrm C)}} &=& 
       - 6 \* {1 \over \beta} \* \z2
       + {7 \over 30}
       + {9 \over 5} \* \z3
       - {199 \over 100} \* \z2
       + {28 \over 5} \* \z2 \* \lntwo
       + {16 \over 5} \* \z2 \* \lnbeta
       + \I \* \pi \* {1 \over \beta} \* (
            3
          - 4 \* \lntwo
          - 2 \* \lnbeta
          )
       - \I \* \pi \* {8 \over 5} \* \z2        
\, ,
\\
\label{eq:diaQV1}
{{\cal F}_1^{(\mathrm{QV})}} &=&
  {1\over \ep^2} \* \biggl\{
          {9 \over 8}
       - \I \* \pi \* {1 \over 2} \* {1 \over \beta}
          \biggr\}
  + {1\over \ep} \* \biggl\{
          {27 \over 16}
       - \I \* \pi \* {1 \over \beta} \* \biggl(
            {3 \over 4}
          + {1 \over 2} \* \lntwo
          \biggr)
          \biggr\}
       + {1 \over \beta} \*  \biggl(
          - {15 \over 2} \* \z2
          + 27 \* \z2 \* \lntwo
          + 18 \* \z2 \* \lnbeta
          \biggr)
          - {473 \over 96}
\\
& &\mbox{}
          - {5081 \over 240} \* \z2
          + 2 \* \z3
          - {64 \over 5} \* \z2 \* \lntwo        
       + \I  \*  \pi   \*  {1 \over \beta}  \*  \biggl(
          - {1 \over 4}
          - {9 \over 2} \* \lntwo
          + {13 \over 2} \* \lntwos
          - 2 \* \z2
          - {5 \over 2} \* \lnbeta
          + 9 \* \lntwo \* \lnbeta
          + 3 \* \lnbetas
          \biggr)
\, ,
\nonumber\\
\label{eq:diaQV2}
{{\cal F}_2^{(\mathrm{QV})}} &=& 
    {1\over \ep} \* \biggl\{
       - {1 \over 2}
       + \I \* \pi \* {1 \over 4} \* {1 \over \beta}
          \biggr\}
       + {9 \over 2} \* {1 \over \beta} \* \z2
       - 3
       - {487 \over 120} \* \z2
       + \z3
       + {44 \over 5} \* \z2 \* \lntwo        
       - \I \* \pi \* {1 \over \beta}  \* \biggl(
            {3 \over 4}
          - {5 \over 2} \* \lntwo
          - {3 \over 2} \* \lnbeta
          \biggr)
\, ,
\\
\label{eq:diaS1}
{{\cal F}_1^{(\mathrm S)}} &=&
  {1\over \ep^2} \* \biggl\{
       - {11 \over 8}
       + \I \* \pi \* {3 \over 8} \* {1 \over \beta^3}
       + \I \* \pi \* {7 \over 8} \* {1 \over \beta}
          \biggr\}
  + {1\over \ep} \* \biggl\{
       - {9 \over 4} \* {1 \over \beta^3} \* \z2
       - {9 \over 4} \* {1 \over \beta} \* \z2
          - {161 \over 48}
       + \I \* \pi \* {1 \over \beta^3}  \*  \biggl(
            {5 \over 4}
          - {3 \over 4} \* \lntwo
          - {3 \over 4} \* \lnbeta
          \biggr)
\\
& &\mbox{}
       + \I \* \pi \* {1 \over \beta}  \*  \biggl(
            {7 \over 8}
          - {1 \over 4} \* \lntwo
          - {3 \over 4} \* \lnbeta
          \biggr)
          \biggr\}
       + {1 \over \beta^3}  \*  \biggl(
          - {15 \over 2} \* \z2
          + {9 \over 2} \* \z2 \* \lntwo
          + {9 \over 2} \* \z2 \* \lnbeta
          \biggr)
       + {1 \over \beta}  \*  \biggl(
            {27 \over 4} \* \z2
          - {45 \over 2} \* \z2 \* \lntwo
          - {27 \over 2} \* \z2 \* \lnbeta
          \biggr)
\nonumber\\
& &\mbox{}
          - {265 \over 288}
          + {5951 \over 240} \* \z2
       + \I \* \pi \* {1 \over \beta^3}  \*  \biggl(
            {17 \over 4}
          - {5 \over 2} \* \lntwo
          + {3 \over 4} \* \lntwos
          - {3 \over 4} \* \z2
          - {5 \over 2} \* \lnbeta
          + {3 \over 2} \* \lntwo \* \lnbeta
          + {3 \over 4} \* \lnbetas
          \biggr)
\nonumber\\
& &\mbox{}
       + \I \* \pi \* {1 \over \beta}  \*  \biggl(
          - {1 \over 2}
          + {17 \over 4} \* \lntwo
          - {23 \over 4} \* \lntwos
          + {5 \over 4} \* \z2
          + {9 \over 4} \* \lnbeta
          - {15 \over 2} \* \lntwo \* \lnbeta
          - {9 \over 4} \* \lnbetas
          \biggr)
\, ,
\nonumber\\
\label{eq:diaS2}
{{\cal F}_2^{(\mathrm S)}} &=& 
    {1\over \ep^2} \* \biggl\{
          1
       - \I \* \pi \* {3 \over 8} \* {1 \over \beta^3}
          \biggr\}
  + {1\over \ep} \* \biggl\{
         {9 \over 4} \* {1 \over \beta^3} \* \z2
       + {2 \over 3}
       + \I \* \pi \* {1 \over \beta^3}  \*  \biggl(
          - 2
          + {3 \over 4} \* \lntwo
          + {3 \over 4} \* \lnbeta
          \biggr)
       + \I \* \pi \* {13 \over 8} \* {1 \over \beta} 
          \biggr\}
\\
& &\mbox{}
       + {1 \over \beta^3}  \*  \biggl(
            12 \* \z2
          - {9 \over 2} \* \z2 \* \lntwo
          - {9 \over 2} \* \z2 \* \lnbeta
          \biggr)
       - {63 \over 4} \* {1 \over \beta} \* \z2  
       + {77 \over 9}
       - {503 \over 120} \* \z2
\nonumber\\
& &\mbox{}
       + \I \* \pi \* {1 \over \beta^3}  \*  \biggl(
          - {21 \over 4}
          + 4 \* \lntwo
          - {3 \over 4} \* \lntwos
          + {3 \over 4} \* \z2
          + 4 \* \lnbeta
          - {3 \over 2} \* \lntwo \* \lnbeta
          - {3 \over 4} \* \lnbetas
          \biggr)
       + \I \* \pi \* {1 \over \beta}  \*  \biggl(
            {27 \over 4}
          - {25 \over 4} \* \lntwo
          - {21 \over 4} \* \lnbeta
          \biggr)
\, ,
\nonumber\\
\label{eq:diaQL1}
{{\cal F}_1^{(\mathrm{QL})}} &=&
  - {1\over \ep^2} \* \biggl\{
         {1 \over 2}
       - \I \* \pi \* {1 \over 6} \* {1 \over \beta} 
       \biggr\}
  + {1\over \ep} \* \biggl\{
        - {1 \over \beta} \* \z2
        + {5 \over 24}
       + \I \* \pi \* {1 \over \beta} \*  \biggl(
            {1 \over 6}
          - {1 \over 3} \* \lntwo
          - {1 \over 3} \* \lnbeta
          \biggr)
       \biggr\}
       - {1 \over \beta} \*  (
            \z2
          - 2 \* \z2 \* \lntwo
          - 2 \* \z2 \* \lnbeta
          )
\\
& &\mbox{}
          - {347 \over 144}
          + {8 \over 15} \* \z2
       + \I \* \pi \* {1 \over \beta} \*  \biggl(
            {2 \over 3}
          - {1 \over 3} \* \lntwo
          + {1 \over 3} \* \lntwos
          - {1 \over 3} \* \z2
          - {1 \over 3} \* \lnbeta
          + {2 \over 3} \* \lntwo \* \lnbeta
          + {1 \over 3} \* \lnbetas
          )
\nonumber\\
& &\mbox{}
+ \nl \* \biggl[
  {1\over \ep^2} \* \biggl\{
       - {1 \over 4}
       + \I \* \pi \* {1 \over 12} \* {1 \over \beta}
       \biggr\}
  - {1\over \ep} \* \biggl\{
         {1 \over \beta} \* \z2
       + {1 \over 8}
       - \I \* \pi \* {1 \over \beta} \*  \biggl(
            {11 \over 36}
          - {1 \over 3} \* \lntwo
          - {1 \over 3} \* \lnbeta
          \biggr)
       \biggr\}
       - {1 \over \beta} \*  \biggl(
            {11 \over 3} \* \z2
          - 4 \* \z2 \* \lntwo
          - 4 \* \z2 \* \lnbeta
          \biggr)
\nonumber\\
& &\mbox{}
          - {185 \over 48}
          - \z2
       + \I \* \pi \* {1 \over \beta} \*  \biggl(
            {175 \over 108}
          - {11 \over 9} \* \lntwo
          + {2 \over 3} \* \lntwos
          - {1 \over 3} \* \z2
          - {11 \over 9} \* \lnbeta
          + {4 \over 3} \* \lntwo \* \lnbeta
          + {2 \over 3} \* \lnbetas
          \biggr)
\biggr]
\, ,
\nonumber\\
\label{eq:diaQL2}
{{\cal F}_2^{(\mathrm{QL})}} &=& 
    {1\over \ep} \* \biggl\{
         {1 \over 3}
       - \I \* \pi \* {1 \over 6} \* {1 \over \beta}
          \biggr\}
       + {1 \over \beta} \* \z2
       - {11 \over 18}
       + {4 \over 5} \* \z2
       + \I \* \pi \* {1 \over \beta} \*  \biggl(
          - {2 \over 3}
          + {1 \over 3} \* \lntwo
          + {1 \over 3} \* \lnbeta
          \biggr)
\\
& &\mbox{}
+ \nl \* \biggl[
    {1\over \ep} \* \biggl\{
         {1 \over 3}
       - \I \* \pi \* {1 \over 6} \* {1 \over \beta}
          \biggr\}
       + 2 \* {1 \over \beta} \* \z2
       + \I \* \pi \* {1 \over \beta} \* \biggl(
          - {49 \over 36}
          + {2 \over 3} \* \lntwo
          + {2 \over 3} \* \lnbeta
          \biggr)
       + {25 \over 18}
\biggr]
\, ,
\nonumber\\
\label{eq:diaGL1}
{{\cal F}_1^{(\mathrm{GL})}} &=&
    {1\over \ep^2} \* \biggl\{
            {5 \over 8}
       - \I \* \pi \* {5 \over 24} \*{1 \over \beta}
          \biggr\}
    + {1\over \ep} \* \biggl\{
         {5 \over 2} \*{1 \over \beta} \* \z2
       + {9 \over 16}
       - \I \* \pi \* {1 \over \beta}  \*  \biggl(
            {61 \over 72}
          - {5 \over 6} \* \lntwo
          - {5 \over 6} \* \lnbeta
          \biggr)
          \biggr\}
\\
& &\mbox{}
       + {1 \over \beta}  \*  \biggl(
            {61 \over 6} \* \z2
          - 10 \* \z2 \* \lntwo
          - 10 \* \z2 \* \lnbeta
          \biggr)
          + {961 \over 96}
          + {5 \over 2} \* \z2
\nonumber\\
& &\mbox{}
       + \I \* \pi \* {1 \over \beta}  \*  \biggl(
          - {959 \over 216}
          + {61 \over 18} \* \lntwo
          - {5 \over 3} \* \lntwos
          + {5 \over 6} \* \z2
          + {61 \over 18} \* \lnbeta
          - {10 \over 3} \* \lntwo \* \lnbeta
          - {5 \over 3} \* \lnbetas
          \biggr)
\, ,
\nonumber\\
\label{eq:diaGL2}
{{\cal F}_2^{(\mathrm{GL})}} &=& 
   - {1\over \ep} \* \biggl\{
         {5 \over 6}
       - \I \* \pi \* {5 \over 12} \* {1 \over \beta}
          \biggr\}
       - 5 \* {1 \over \beta} \* \z2
       - {137 \over 36}
       + \I \* \pi \* {1 \over \beta} \*  \biggl(
            {257 \over 72}
          - {5 \over 3} \* \lntwo
          - {5 \over 3} \* \lnbeta
          \biggr)
\, ,
\\
\label{eq:diaGV1}
{{\cal F}_1^{(\mathrm{GV})}} &=&
    {1\over \ep^2} \* \biggl\{
          {15 \over 16}
       - \I \* \pi \* {3 \over 8} \* {1 \over \beta}
          \biggr\}
  + {1\over \ep} \* \biggl\{
         {9 \over 4} \* {1 \over \beta} \* \z2
       + {19 \over 32}
       + \I \* \pi \* {1 \over \beta}  \*  \biggl(
          - {7 \over 8}
          + {3 \over 4} \* \lntwo
          + {3 \over 4} \* \lnbeta
          \biggr)
          \biggr\}
       + {1 \over \beta}  \*  \biggl(
            6 \* \z2
          - {9 \over 2} \* \z2 \* \lntwo
          - {9 \over 2} \* \z2 \* \lnbeta
          \biggr)
\\
& &\mbox{}
          + {5887 \over 960}
          - {49 \over 40} \* \z3
          - {1901 \over 300} \* \z2
          - {21 \over 10} \* \z2 \* \lntwo
          - {7 \over 5} \* \z2 \* \lnbeta
\nonumber\\
& &\mbox{}
       + \I \* \pi \* {1 \over \beta}  \*  \biggl(
          - {45 \over 16}
          + 2 \* \lntwo
          - {3 \over 4} \* \lntwos
          + {3 \over 4} \* \z2
          + 2 \* \lnbeta
          - {3 \over 2} \* \lntwo \* \lnbeta
          - {3 \over 4} \* \lnbetas
          \biggr)
       + \I \* \pi  \* {7 \over10} \* \z2
\, ,
\nonumber\\
\label{eq:diaGV2}
{{\cal F}_2^{(\mathrm{GV})}} &=& 
    - {1\over \ep} \* \biggl\{
         {3 \over 4}
       - \I \* \pi \* {3 \over 8} \* {1 \over \beta}
          \biggr\}
       - 3 \* {1 \over \beta} \* \z2
       - {289 \over 120}
       + {244 \over 75} \* \z2
       - {7 \over 5} \* \z3
       - {12 \over 5} \* \z2 \* \lntwo
       - {8 \over 5} \* \z2 \* \lnbeta
\\
& &\mbox{}
       + \I \* \pi \* {1 \over \beta} \* \biggl(
            {29 \over 16}
          - \lntwo
          - \lnbeta
          \biggr)
       + \I \* \pi  \* {4 \over 5} \* \z2        
\, .
\nonumber
\end{eqnarray}
}
Due to the time-like kinematics we naturally encounter imaginary parts.
We should point out again, that the diagram $QL$ contains both $\nl$ light
quark loops and one heavy quark loop, while $GL$ is the gluon and the ghost loop contribution 
to the gluon self-energy.

While our calculation~(\ref{eq:addFdiag}) has been performed for a color-singlet (${\bf 1}$) current coupling 
within the color-SU$(N_c)$ gauge group, nevertheless the results of Eqs.~(\ref{eq:diaL1})--(\ref{eq:diaGV2}) 
can also be applied to considerations of color-octet (${\bf 8}$) currents.
In the latter case the relevant color factors are listed in column ${\bf 8}$ of Tab.~\ref{tab:FFcolor}. 
Such decompositions are needed, for instance, in the description of heavy quarkonium states 
(see e.g.~\cite{Petrelli:1997ge,Hagiwara:2008df,Kiyo:2008bv}).
At one-loop, i.e. for the diagrams $OneV$ and $OneS$ 
in Fig.~\ref{fig:massiveFF1} the transition from a quark-pair in a color-singlet
state to one in a color-octet state amounts to the simple replacement of $C_F \to (C_F-{C_A/2})$.
At two loops, this is not true for the individual Feynman diagrams, see Tab.~\ref{tab:FFcolor}.
However, for the complete two-loop Coulomb corrections again the same simple
replacement rule holds, as we want to illustrate next.

To that end, we present our results in the form as they enter in physical processes,
\begin{eqnarray}
  \label{eq:Xqqb18}
  X &\to& \left( q{\bar q} \right)_{\bf I}\, , 
  \qquad\qquad
  {\bf I = 1,8}\, ,
\qquad
\end{eqnarray}
where $(q{\bar q})_{\bf I}$ represents a (heavy) color-singlet or octet final state.
Near threshold, considering only Coulomb enhanced contributions, the relevant Born cross section ${\sigma}^{(0)}_{\bf I}$
factorizes, so that
\begin{eqnarray}
  \label{eq:sigmaCoul}
  {\sigma}_{X \to \left( q{\bar q} \right)_{\bf I}} &=&
  {\sigma}^{(0)}_{\bf I} \, 
  \left(
    1 
    + a_s  \, \Delta_{\bf I}^{(1)}
    + a_s^2\, \Delta_{\bf I}^{(2)}
  \right)
  \, , 
\qquad
\end{eqnarray}
where the $\Delta_{\bf I}^{(i)}$ can be expressed in terms of the 
threshold expansions of the renormalized form factors above threshold, cf. Eq.~(\ref{eq:tmlkdef}).
For the color-singlet case $(q{\bar q})_{\bf 1}$ we have from Eq.~(\ref{eq:addFdiag}),
\begin{eqnarray}
  \label{eq:Delta-def}
  \Delta_{\bf 1}^{(1)} &=& 
  2 \Bigl( \Re {\cal F}_1^{(1)}
  + \Re {\cal F}_2^{(1)} 
  \Bigr) 
\, , \\
  \Delta_{\bf 1}^{(2)} &=& 
  \Bigl( \Re {\cal F}_1^{(1)}
  + \Re {\cal F}_2^{(1)} 
  \Bigr)^2
  + 
  \Bigl( \Im {\cal F}_1^{(1)}
  + \Im {\cal F}_2^{(1)} 
  \Bigr)^2
  +
  2 \Bigl( \Re {\cal F}_1^{(2)}
  + \Re {\cal F}_2^{(2)} 
  \Bigr)
  \, .
\end{eqnarray}
Upon adding Eqs.~(\ref{eq:diaL1})--(\ref{eq:diaGV2}) with the appropriate color
coefficients from Tab.~\ref{tab:FFcolor} and performing the necessary renormalization, 
i.e. the threshold expansion of the time-like counter-terms~(\ref{eq:F1mctl1})--(\ref{eq:F2mctl1}),
we find the Coulomb corrections for the color-singlet final state as 
\begin{eqnarray}
  \label{eq:DeltaCoul1}
  \Delta_{\bf 1}^{(1)} &=&
  2 \* \cf \* {\pi^2 \over \beta}
\, ,
\\
  \label{eq:DeltaCoul2}
  \Delta_{\bf 1}^{(2)} &=&
  2\* \cf \* \Biggl\{
  \left(
          {31 \over 9}\*\ca 
        - 16\*\cf 
        - {10 \over 9}\*\nl 
        - 2\*\b0 \* \lntwo 
        - 2\*\b0 \* \lnbeta
        \right) \* {\pi^2 \over \beta}
+
  {2 \over 3} \* \cf \* {\pi^4 \over \beta^2}
  \Biggr\}
\, .
\qquad
\end{eqnarray}
Note, that all divergencies in $\ep$ as well as all higher inverse powers in
$\beta$ have canceled in Eqs.~(\ref{eq:DeltaCoul1}) and
(\ref{eq:DeltaCoul2}), so that we retain at most $(a_s/\beta)^L$ at $L$-loops.
For the octet final state $(q{\bar q})_{\bf 8}$ we can evaluate Eq.~(\ref{eq:addFdiag}) with 
the color factors appropriately replaced according to Tab.~\ref{tab:FFcolor}.
The resulting expressions for $\Delta_{\bf 8}^{(1)}$ and $\Delta_{\bf 8}^{(2)}$ after these modifications 
are then indeed obtained as in Eqs.~(\ref{eq:DeltaCoul1}), (\ref{eq:DeltaCoul2}) 
with the replacement of the color factor $2\cf \to (2\cf - \ca)$. 
This is in complete agreement 
with results from potential non-relativistic QCD (e.g. \cite{Czarnecki:1997vz,Pineda:2006ri}).

%
%
%
\section{Applications}
\label{sec:appl}
%
%

In this Section we present several applications of the new
results for the heavy-quark form factor. 
Particular emphasis will be put on new three-loop predictions, because 
the order $\epsilon$ results at two loops are necessary ingredients for 
a full analytic three-loop calculation.
The applications will mostly be concerned with the electric form factor ${\cal F}_1$. 
We discuss the property of exponentiation as well as its role in the
determination of massive $n$-parton gauge amplitudes in the high-energy limit 
from massless ones.

\subsection{Exponentiation of form factor}
Let us start by recalling briefly the exponentiation for the heavy-quark form factor.
This feature is based on the universality of soft and collinear radiation 
and the respective singular terms in ${\cal F}_1$, 
i.e. the poles in $\ep$ and the large logarithms $\ln(m)$ of Sudakov type~\cite{Sudakov:1954sw},
so that ${\cal F}_1$ fulfills the following evolution equation (see e.g.~\cite{Collins:1980ih,Mitov:2006xs}),
\begin{eqnarray}
  \label{eq:ffdeq}
- \mu^2 {\partial \over \partial \mu^2} \ln {\cal F}_1\left({Q^2 \over \mu^2},{m^2 \over \mu^2},\alpha_{s},\ep\right) 
&=&
  {1 \over 2} \: K\left({m^2 \over \mu^2},\alpha_{s},\ep \right)
+ {1 \over 2} \: G\left({Q^2 \over \mu^2},\alpha_{s},\ep \right) \, ,
\end{eqnarray}
where the dependence on the various scales has been made explicit. 
QCD factorization at the scale $\mu$ 
allows to separate in Eq.~(\ref{eq:ffdeq}) 
the dependence on the hard scale $Q$ from that on the heavy-quark mass $m$.
To logarithmic accuracy, the former rests entirely in the function $G$ while
the latter is associated with the function $K$.
Both functions, $G$ and $K$, are subject to renormalization
group equations~\cite{Collins:1980ih,Collins:1989bt} governed by the same 
(well-known) cusp anomalous dimension $A$~\cite{Kodaira:1982nh,Moch:2004pa,Vogt:2004mw},
\begin{eqnarray}
  \label{eq:KGdeq}
  - \lim_{m \to 0} \mu^2 {d \over d \mu^2} K\left({m^2 \over \mu^2},\alpha_{s},\ep \right) 
  \, = \,
  \mu^2 {d \over d \mu^2} G\left({Q^2 \over \mu^2},\alpha_{s},\ep \right) 
  \, = \,  A(\alpha_{s}) \, .
\end{eqnarray}
In dimensional regularization, 
the solution of the evolution equation~(\ref{eq:ffdeq}) 
proceeds in complete analogy to the construction for the form factors 
of massless quarks and gluons, see for example~\cite{Moch:2005id,Magnea:1990zb,Moch:2005tm,Dixon:2008gr}.
In the massive case, the necessary integration constant is parameterized through the matching function $C$.
For the (ultraviolet) renormalized massive form factor ${\cal F}_1$ 
with space-like virtuality $q^2 = - Q^2 <0$ and
in terms of the renormalized coupling $\alpha_{s}(\mu^2)$ we arrive at 
\begin{eqnarray}
\label{eq:massFFexp}
{\lefteqn{
    {\cal F}_1 \left({Q^2 \over \mu^2},{m^2 \over \mu^2},a_{s}(\mu^2),\ep \right)
    \, = \,
    C({\bar a}(\mu^2,\ep),\ep) \,
     }}
\\ &&
\times \exp\,
\left( \,
  {1 \over 2}\, 
  \int\limits_0^{Q^2/\mu^2}\, {d \xi \over \xi}
  \Bigg\{
  G(\bar a(\xi \mu^2,\ep)) 
  +
  K(\bar a(\xi \mu^2 m^2/Q^2,\ep)) 
  -
  \int\limits_{\xi m^2/Q^2}^\xi\, {d \lambda \over \lambda}\
  A({\bar a}(\lambda \mu^2,\ep))
  \Bigg\}
\right)
\, , \nonumber
\end{eqnarray}
where all quantities on the right hand side are defined in $D$ dimensions 
(note a change of sign in the definition of $K$ compared to Ref.~\cite{Mitov:2006xs}).
They are functions of the $D$-dimensional strong coupling ${\bar a}$ 
and can be expressed in terms of the ordinary coupling $\alpha_{s}$ in four dimensions
(see e.g.~\cite{Moch:2005id}) 
like the functions $C({\bar a})$, $G({\bar a})$ and $K({\bar a})$.

The coefficients in the perturbative expansion of the cusp anomalous dimension $A$ 
and the function $G$ agree with those of the form factor
for massless quarks, see e.g.~\cite{Moch:2005id,Moch:2005tm}.
Due to the heavy-quark mass acting as an additional regulator 
in the collinear limit the coefficients of the infrared counter term $K$ 
and also the matching function $C$ take, on the other hand, particular values.
Their respective coefficients read 
{\small
\begin{eqnarray}
\label{eq:k1}
  K^{(1)} & = &
           - 2 \* \cf
\, ,
\\
\label{eq:k2}
  K^{(2)} & = &
           \cf^2  \*  (
          - 3
          + 24 \* \z2
          - 48 \* \z3
          )
       - \cf \* \ca  \*  \biggl(
            {373 \over 27}
          + 30 \* \z2
          - 60 \* \z3
          \biggr)
       + \nl \* \cf  \*  \biggl(
            {10 \over 27}
          + 4 \* \z2
          \biggr)
\, ,
\\
\label{eq:c1}
  C^{(1)} & = &
         \cf \* (4 + \z2)
       + \ep \* \cf \* \biggl(
            8
          + {1 \over 2} \* \z2
          - {2 \over 3} \* \z3
          \biggr)
       + \ep^2 \* \cf \* \biggl(
            16
          + 2 \* \z2
          - {1 \over 3} \* \z3
          + {9 \over 20} \* \z2^2
          \biggr)
+ \calO\left(\ep^3\right)
\, ,
\quad
\\
\label{eq:c2}
  C^{(2)} & = &
         \cf^2 \* \biggl\{
            {241\over 8} 
          + 26\*\z2 
          - 6\*\z3 
          - 48\*\z2\*\lntwo 
          - {163\over 10}\*\z2^2 
       + \ep \* \biggl(
          - {557\over 16} 
          + {399\over 4}\*\z2 
          + {286\over 3}\*\z3 
          - 24\*\z2\*\lntwo 
\\
& &\mbox{}
          + 192\*\a4 
          + 8\*\lntwof 
          + 96\*\z2\*\lntwos 
          - {1363\over 10}\*\z2^2 
          - {146\over 3}\*\z2\*\z3 
          - 6\*\z5 
          \biggr)
          \biggr\}
\nonumber\\
&+&\mbox{}
         \ca \* \cf \* \biggl\{
            {12877\over 648} 
          + {323\over 18}\*\z2 
          + {89\over 9}\*\z3 
          + 24\*\z2\*\lntwo 
          - {47\over 5}\*\z2^2 
       + \ep \* \biggl(
            {629821\over 3888} 
          + {7805\over 108}\*\z2 
          - {1162\over 27}\*\z3 
          + {209\over 2}\*\z2^2
\nonumber\\
& &\mbox{} 
          - 96\*\a4 
          - 4\*\lntwof 
          + 12\*\z2\*\lntwo 
          - 48\*\z2\*\lntwos 
          + {83\over 3}\*\z2\*\z3 
          - 208\*\z5 
          \biggr)
          \biggr\}
\nonumber\\
&+&\mbox{}
         \nl \* \cf \* \biggl\{
          - {1541\over 324} 
          - {37\over 9}\*\z2
          - {26\over 9}\*\z3 
       + \ep \* \biggl(
          - {46205\over 1944}
          - {673\over 54}\*\z2 
          - {182\over 27}\*\z3 
          - {49\over 5}\*\z2^2
          \biggr)
          \biggr\}
+ \calO\left(\ep^2\right)
\nonumber
\nonumber\, .
\end{eqnarray}
}
Eqs.~(\ref{eq:k1})--(\ref{eq:c1}) have been obtained before~\cite{Mitov:2006xs}. 
The ${\cal O}(\epsilon)$ term of $C^{(2)}$ in Eq.~(\ref{eq:c2}) is a new result.
Note also that we have changed the definition of $C^{(2)}$ compared to Ref.~\cite{Mitov:2006xs}, 
because matching at two loops has now been extended to ${\cal O}(\epsilon)$ consistently.

With these values for $A$, $G$, $K$ and $C$ and, 
upon performing the integrations over $\xi$ and $\lambda$ in Eq.~(\ref{eq:massFFexp}) 
the exponential for ${\cal F}_1$ generates all large logarithms 
in the heavy-quark mass $m$ (even including higher orders of $\ep$).
In particular, we are able to reproduce the high-energy expansions 
through two loops in Eqs.~(\ref{eq:F1rl1-x0}) and (\ref{eq:F1rl2-x0}).
This result provides an independent check of our result, 
because the exponentiation~(\ref{eq:massFFexp}) relies on functions 
$A$ and $G$ that have been derived in massless computations.

Let us at this point make two remarks on the exponential~(\ref{eq:massFFexp}).
First, the derivation of Eq.~(\ref{eq:massFFexp}) relies on the heavy-quark masses
(acting as infrared regulator) to be attached to external lines.
By construction, we have omitted all contributions from 
diagram $QL$ in Fig.~\ref{fig:massiveFF2} with a virtual heavy-quark in the
loop in Eq.~(\ref{eq:massFFexp}) and the comparison to Eqs.~(\ref{eq:F1rl1-x0}), (\ref{eq:F1rl2-x0}).
These contributions are finite after performing the ultraviolet renormalization, 
but they still do contain Sudakov logarithms $L$, see Eq.~(\ref{eq:L-def}).
It is well-known, that these remaining logarithms also obey 
an exponentiation similar to Eq.~(\ref{eq:massFFexp}), see e.g. Ref.~\cite{Mitov:2006xs} 
and related discussions in electro-weak theories~\cite{Kuhn:2001hz,Feucht:2003yx}.
Here, we do not elaborate on this point further.

Second, in the exponential expression Eq.~(\ref{eq:massFFexp}) for the form factor
we have used the standard $\MSbar$ coupling running with $\nl$ light flavors.
In order to compare Eq.~(\ref{eq:massFFexp}) 
or rather its expanded version to the fixed-order results Eqs.~(\ref{eq:F1rl1}) and (\ref{eq:F1rl2}) 
which also employ the $\MSbar$-scheme, but a running coupling with a total number of flavors $\nf=\nl+1$,
one has to apply the decoupling relations~\cite{Appelquist:1974tg}.
The necessary decoupling constant for $\alpha_{s}$ at flavor thresholds is known to
$\calO(\alpha_{s}^3)$~\cite{Larin:1994va,Chetyrkin:1997un}
(see also Refs.~\cite{Schroder:2005hy,Chetyrkin:2005ia} for $\calO(\alpha_{s}^4)$).
To relate the two results, we use the following relation for $\alpha_{s}$,
\begin{eqnarray}
  \label{eq:alphas-dec-ep}
  \alpha_{s}^{(\nl)} & = &
    \asnf
  + \biggl(\asnf \biggr)^2 \* \biggl\{
          - {2 \over 3} \* L_{\mu}
       - \ep  \*  \biggl(
            {1 \over 3} \* L_{\mu}^2
          + {1 \over 3} \* \z2
          \biggr)
       - \ep^2  \* \biggl(
            {1 \over 9} \* L_{\mu}^3
          + {1 \over 3} \* \z2 \* L_{\mu}
          - {2 \over 9} \* \z3
          \biggr)
  \biggr\}
\\
& &\mbox{} 
  + \biggl(\asnf\biggr)^3  \*  \biggl\{
            {4 \over 9} \* L_{\mu}^2
          - {10 \over 3} \* \ca \* L_{\mu}
          - 2 \* \cf \* L_{\mu}
          + {16 \over 9} \* \ca
          - {15 \over 2} \* \cf
\nonumber\\
& &\mbox{} 
       + \ep  \*  \biggl(
            {4 \over 9} \* L_{\mu}^3
          - {10 \over 3} \* \ca \* L_{\mu}^2
          + {32 \over 9} \* \ca \* L_{\mu}
          - {29 \over 3} \* \cf \* L_{\mu}
          + {4 \over 9} \* \z2 \* L_{\mu}
          - {86 \over 27} \* \ca
          - {5 \over 3} \* \ca \* \z2
          + {35 \over 12} \* \cf
          + \cf \* \z2
          \biggr)
\nonumber\\
& &\mbox{} 
       - \ep^2  \*  \biggl(
            2 \* \cf \* L_{\mu}^3
          + {8 \over 3} \* \cf \* L_{\mu}^2
          + 2 \* \cf \* \z2 \* L_{\mu}
          + {8 \over 3} \* \cf \* \z2
          \biggr)
  \biggr\}
       + \calO(\ep^3) 
  + \calO \biggl( \biggl(\alpha_{s}^{(\nf)}\biggr)^{\,4} \biggr) \, ,
\nonumber
\end{eqnarray}
where $\alpha_{s}^{(\nl)}$ is the standard $\MSbar$ coupling for $\nl$ 
quark flavors expanded in terms of $\alpha_{s}^{(\nf)}$ for $\nf=\nl+1$ flavors, 
both evaluated at the scale $\mu^2$, and
\begin{eqnarray}
 \label{eq:Lmu-def}
 L_{\mu} = \ln\left( {\mu^2\over m^2}\right)\, .
 \end{eqnarray}
Eq.~(\ref{eq:alphas-dec-ep}) uses the pole-mass $m$ 
and the higher order terms in $\epsilon$ in Eq.~(\ref{eq:alphas-dec-ep}) are needed 
for the higher order expansion coefficients in Eqs.~(\ref{eq:F1rl1-x0}) and (\ref{eq:F1rl2-x0}).

Now we are ready to employ Eq.~(\ref{eq:massFFexp}) to obtain 
a partial new three-loop prediction for ${\cal F}_1$ in the limit $Q^2 \gg m^2$.
With the necessary expansion coefficients for $A$ and $G$ to third order, 
we obtain at the scale $\mu = m$,
{\small
\begin{eqnarray}
\label{eq:F1rl3-x0}
{\cal F}_1^{(3)} &=&
         \cf^3 \* \biggl\{
         {1 \over \ep^3} \* \biggl(
            {4 \over 3} \* L^3
          - 4 \* L^2
          + 4 \* L
          - {4 \over 3}
          \biggr)
       + {1 \over \ep^2} \* (
          - 2 \* L^4
          + 10 \* L^3
          - (
            22
          - 4 \* \z2 ) \* L^2
          + (
            22
          - 8 \* \z2 ) \* L
          - 8
          + 4 \* \z2
          )
\\
& &\mbox{}
       + {1 \over \ep} \* \biggl(
            {5 \over 3} \* L^5
          - {34 \over 3} \* L^4
          + \biggl(
            {137 \over 3}
          - 6 \* \z2
          \biggr) \* L^3
          - \biggl(
            89
          - 56 \* \z3
          \biggr) \* L^2
          + \biggl(
            129
          + 88 \* \z2
          - 136 \* \z3
          - 96 \* \z2 \* \lntwo
          - {236 \over 5} \* \z2^2
          \biggr) \* L
\nonumber\\
& &\mbox{}
          - {485 \over 6}
          - 88 \* \z2
          + 96 \* \z2 \* \lntwo
          + {172 \over 3} \* \z3
          + 28 \* \z2^2
          + {32 \over 3} \* \z2 \* \z3
          + 80 \* \z5
          \biggr)
          - L^6
          + {17 \over 2} \* L^5
          - \biggl(
            {148 \over 3}
          - {16 \over 3} \* \z2
          \biggr) \* L^4
\nonumber\\
& &\mbox{}
          + \biggl(
            {494 \over 3}
          + {17 \over 3} \* \z2
          - {268 \over 3} \* \z3
          \biggr) \* L^3
          + \biggl(
          - 332
          - 123 \* \z2
          + 48 \* \z2 \* \lntwo
          + 340 \* \z3
          + {302 \over 5} \* \z2^2
          \biggr) \*  L^2
          + \biggl(
            250
          + 585 \* \z2
\nonumber\\
& &\mbox{}
          - 192 \* \z2 \* \lntwo
          - 520 \* \z3
          + 192 \* \z2 \* \lntwos
          + 16 \* \lntwof
          + 384 \* \a4
          - {1942 \over 5} \* \z2^2
          - 8 \* \z2 \* \z3
          - 276 \* \z5
          \biggr) \* L
       \biggr\}
\nonumber\\
&+&\mbox{}
         \cf^2 \* \ca \* \biggl\{
         {1 \over \ep^3} \* \biggl(
          - {22 \over 3} \* L^2
          + {44 \over 3} \* L
          - {22 \over 3}
          \biggr)
       + {1 \over \ep^2} \* \biggl(
            {11 \over 3} \* L^3
          + \biggl(
            {2 \over 9}
          - 4 \* \z2
          \biggr) \* L^2
          - \biggl(
            {1 \over 9}
          - {2 \over 3} \* \z2
          + 4 \* \z3
          \biggr) \* L
          - {34 \over 9}
          + {10 \over 3} \* \z2
          + 4 \* \z3
          \biggr)
\nonumber\\
& &\mbox{}
       + {1 \over \ep} \* \biggl(
            {11 \over 9} \* L^4
          - \biggl(
            {523 \over 18}
          - 6 \* \z2
          \biggr) \* L^3
          + \biggl(
            {6107 \over 54}
          + {19 \over 3} \* \z2
          - 50 \* \z3
          \biggr) \* L^2
          - \biggl(
            {5396 \over 27}
          - {5 \over 3} \* \z2
          - {362 \over 3} \* \z3
          - 48 \* \z2 \* \lntwo
\nonumber\\
& &\mbox{}
          + {26 \over 5} \* \z2^2
          \biggr) \* L
          + {10585 \over 108}
          + {320 \over 9} \* \z2
          - 48 \* \z2 \* \lntwo
          - {1444 \over 9} \* \z3
          + {1222 \over 45} \* \z2^2
          - {28 \over 3} \* \z2 \* \z3
          - 40 \* \z5
          \biggr)
          - {11 \over 4} \* L^5
\nonumber\\
& &\mbox{}
          + \biggl(
            {4289 \over 108}
          - {16 \over 3} \* \z2
          \biggr) \* L^4
          - \biggl(
            {6260 \over 27}
          + {97 \over 18} \* \z2
          - {232 \over 3} \* \z3
          \biggr) \* L^3
          + \biggl(
            {64625 \over 81}
          + {266 \over 3} \* \z2
          - 24 \* \z2 \* \lntwo
          - {3743 \over 9} \* \z3
\nonumber\\
& &\mbox{}
          - {143 \over 5} \* \z2^2
          \biggr) \* L^2
          + \biggl(
          - {178337 \over 162}
          - {5959 \over 18} \* \z2
          + 96 \* \z2 \* \lntwo
          + {3706 \over 3} \* \z3
          - 96 \* \z2 \* \lntwos
          - 8 \* \lntwof
          - 192 \* \a4
\nonumber\\
& &\mbox{}
          + {4279 \over 15} \* \z2^2
          - 46 \* \z2 \* \z3
          - 194 \* \z5
          \biggr) \* L
       \biggr\}
\nonumber\\
&+&\mbox{}
         \cf \* \ca^2 \* \biggl\{
         {1 \over \ep^3} \* \biggl(
            {242 \over 27} \* L
          - {242 \over 27}
          \biggr)
       + {1 \over \ep^2} \* \biggl(
          - \biggl(
            {2086 \over 81}
          - {44 \over 9} \* \z2
          \biggr) \* L
          + {1690 \over 81}
          - {44 \over 9} \* \z2
          + {44 \over 9} \* \z3
          \biggr)
       + {1 \over \ep} \* \biggl(
          \biggl(
            {245 \over 9}
          - {536 \over 27} \* \z2
\nonumber\\
& &\mbox{}
          + {44 \over 9} \* \z3
          + {88 \over 15} \* \z2^2
          \biggr) \* L
          - {139345 \over 8748}
          - {7163 \over 243} \* \z2
          + {3526 \over 27} \* \z3
          - {166 \over 15} \* \z2^2
          - {88 \over 9} \* \z2 \* \z3
          - {136 \over 3} \* \z5
          \biggr)
          - {121 \over 54} \* L^4
\nonumber\\
& &\mbox{}
          + \biggl(
            {2869 \over 81}
          - {44 \over 9} \* \z2
          \biggr) \* L^3
          + \biggl(
          - {18682 \over 81}
          + {26 \over 9} \* \z2
          + 88 \* \z3
          - {44 \over 5} \* \z2^2
          \biggr) \*  L^2
          + \biggl(
            {1045955 \over 1458}
          + {17366 \over 81} \* \z2
          - {17464 \over 27} \* \z3
\nonumber\\
& &\mbox{}
          - {94 \over 3} \* \z2^2
          + {88 \over 3} \* \z2 \* \z3
          + 136 \* \z5
          \biggr) \* L
       \biggr\}
\nonumber\\
&+&\mbox{}
         \cf^2 \* \nl \* \biggl\{
         {1 \over \ep^3} \* \biggl(
            {4 \over 3} \* L^2
          - {8 \over 3} \* L
          + {4 \over 3}
          \biggr)
       + {1 \over \ep^2} \* \biggl(
          - {2 \over 3} \* L^3
          + {4 \over 9} \* L^2
          + \biggl(
            {10 \over 9}
          + {4 \over 3} \* \z2
          \biggr) \* L
          - {8 \over 9}
          - {4 \over 3} \* \z2
          \biggr)
       + {1 \over \ep} \* \biggl(
          - {2 \over 9} \* L^4
          + {41 \over 9} \* L^3
\nonumber\\
& &\mbox{}
          - \biggl(
            {481 \over 27}
          + {10 \over 3} \* \z2
          \biggr) \* L^2
          + \biggl(
            {599 \over 27}
          - {2 \over 3} \* \z2
          + {8 \over 3} \* \z3
          \biggr) \* L
          + {553 \over 162}
          + {10 \over 9} \* \z2
          + {328 \over 27} \* \z3
          - {56 \over 9} \* \z2^2
          \biggr)
          + {1 \over 2} \* L^5
          - {355 \over 54} \* L^4
\nonumber\\
& &\mbox{}
          + \biggl(
            {1016 \over 27}
          + {29 \over 9} \* \z2
            \biggr) \* L^3
          + \biggl(
          - {18817 \over 162}
          - {50 \over 3} \* \z2
          + {116 \over 9} \* \z3
            \biggr) \* L^2
          + \biggl(
            {9406 \over 81}
          + {341 \over 9} \* \z2
          - {988 \over 9} \* \z3
          - {196 \over 15} \* \z2^2
            \biggr) \* L
       \biggr\}
\nonumber\\
&+&\mbox{}
         \cf \* \ca \* \nl \* \biggl\{
         {1 \over \ep^3} \* \biggl(
          - {88 \over 27} \* L
          + {88 \over 27}
          \biggr)
       + {1 \over \ep^2} \* \biggl(
            \biggl(
            {668 \over 81}
          - {8 \over 9} \* \z2
          \biggr) \* L
          - {596 \over 81}
          + {8 \over 9} \* \z2
          - {8 \over 9} \* \z3
          \biggr)
       + {1 \over \ep} \* \biggl(
          - \biggl(
            {418 \over 81}
          - {80 \over 27} \* \z2
          + {56 \over 9} \* \z3
          \biggr) \* L
\nonumber\\
& &\mbox{}
          - {8659 \over 2187}
          + {2594 \over 243} \* \z2
          - {964 \over 81} \* \z3
          + {44 \over 15} \* \z2^2
          \biggr)
          + {22 \over 27} \* L^4
          - \biggl(
            {974 \over 81}
          - {8 \over 9} \* \z2
          \biggr) \* L^3
          + \biggl(
            {5876 \over 81}
          + {16 \over 3} \* \z2
          - 8 \* \z3
          \biggr) \* L^2
\nonumber\\
& &\mbox{}
          + \biggl(
          - {154919 \over 729}
          - {5864 \over 81} \* \z2
          + {724 \over 9} \* \z3
          + {44 \over 15} \* \z2^2
          \biggr) \* L
       \biggr\}
\nonumber\\
&+&\mbox{}
         \cf \* \nl^2 \* \biggl\{
         {1 \over \ep^3} \* \biggl(
            {8 \over 27} \* L
          - {8 \over 27}
          \biggr)
       + {1 \over \ep^2} \* \biggl(
          - {40 \over 81} \* L
          + {40 \over 81}
          \biggr)
       + {1 \over \ep} \* \biggl(
          - {8 \over 81} \* L
          + {2417 \over 2187}
          - {20 \over 27} \* \z2
          - {8 \over 81} \* \z3
          \biggr)
       - {2 \over 27} \* L^4
       + {76 \over 81} \* L^3
\nonumber\\
& &\mbox{}
          - \biggl(
            {406 \over 81}
          + {8 \over 9} \* \z2
          \biggr) \* L^2
          + \biggl(
            {9838 \over 729}
          + {152 \over 27} \* \z2
          + {16 \over 27} \* \z3
          \biggr) \* L
       \biggr\}
\nonumber\\
&+&\mbox{}
         \cf^2 \* \biggl\{
         {1 \over \ep} \* \biggl(
          - {4\over 3} \* \z2 \* L^2
          - \biggl(
            15 
          - {8\over 3} \* \z2 
          \biggr) \* L
          + 15
          - {4\over 3} \* \z2
          \biggr)
         + {4 \over 3} \* \z2 \* L^3
         + \biggl(
            {15\over 2}
          - {16\over 3} \* \z2
          + {8\over 9} \* \z3
          \biggr) \* L^2
\nonumber\\
& &\mbox{}
         + \biggl(
          - {50\over 3}
          + {34\over 3} \* \z2
          - {16\over 9} \* \z3
          - {8\over 3} \* \z2^2
          \biggr) \* L
       \biggr\}
\nonumber\\
&+&\mbox{}
         \cf \* \ca \* \biggl\{
         {1 \over \ep} \* \biggl(
           \biggl(
            {32 \over 9}
          + {22\over 9} \* \z2
          \biggr) \* L
          - {32\over 9}
          - {22\over 9} \* \z2
          \biggr)
          - {16 \over 9} \* L^2
          - \biggl(
            {28 \over 27}
          + {224 \over 27} \* \z2
          + {44 \over 27} \* \z3
          - {4 \over 3} \* \z2^2
          \biggr) \* L
       \biggr\}
\nonumber\\
&+&\mbox{}
         \cf \* \nl \* \biggl\{
         {1 \over \ep} \* \biggl(
          - {4 \over 9} \* \z2 \* L
          + {4 \over 9} \* \z2
          \biggr)
         + \biggl(
            {20 \over 27} \* \z2
          + {8 \over 27} \* \z3
          \biggr) \* L
       \biggr\}
\nonumber\\
&-&\mbox{}
         {1 \over {6 \* \ep}} \* K^{(3)}
       + \calO(\ep^0 L^0) 
\nonumber
\, .
\end{eqnarray}
}
As detailed above, we have omitted the finite contributions arising from virtual heavy-flavor
lines, i.e. diagram $QL$ in Fig.~\ref{fig:massiveFF2} with the 
heavy-quark in the loop and higher order generalizations thereof.
All singularities in $\ep$ however are controlled by the decoupling
relation~(\ref{eq:alphas-dec-ep}) 
which also gives rise to the terms proportional to $\cf^2, \cf \ca$ and $\cf\nl$.
The latter are explicitly displayed in Eq.~(\ref{eq:F1rl3-x0}).

The accuracy of the prediction for ${\cal F}_1^{(3)}$ in Eq.~(\ref{eq:F1rl3-x0}) 
is limited by lacking knowledge about the three-loop function $K^{(3)}$ controlling the
infrared sector of the massive theory.
$K^{(3)}$ has been kept explicitly in Eq.~(\ref{eq:F1rl3-x0}) above.
The constant terms of order $\calO(\ep^0 L^0)$ in ${\cal F}_1^{(3)}$
contain the sum of $G^{(3)}$ including its order $\calO (\ep)$-part 
which became available recently~\cite{Baikov:2009bg},
as well as the currently unknown three-loop matching function $C^{(3)}$.
Therefore we have truncated ${\cal F}_1^{(3)}$ as given in Eq.~(\ref{eq:F1rl3-x0}).

\subsection{Massive \emph{n}-parton amplitudes}
Let us again briefly summarize the key features of 
relating loop amplitudes with massive partons to massless ones. 
QCD factorization implies that a massive amplitude ${\cal M}^{(m)}$ 
for any given physical process shares essential properties 
with the corresponding massless amplitude ${\cal M}^{(m=0)}$ 
in the limit when all kinematical invariants are large compared to the
heavy-quark mass $m$.
One typically encounters in ${\cal M}^{(m=0)}$ two types of singularities, 
soft and collinear ones, which are related to the emission of gluons with vanishing energy 
and to collinear parton radiation off massless hard partons, respectively. 
In dimensional regularization, these appear explicitly as
factorizing poles in $\ep$, see~\cite{Catani:1998bh,Sterman:2002qn}, 
and the associated anomalous dimensions are currently subject of active 
studies~\cite{Dixon:2008gr,Aybat:2006wq,Aybat:2006mz,Becher:2009cu,Gardi:2009qi,Dixon:2009gx}. 
In the massive case, the soft singularities remain in ${\cal M}^{(m)}$ as single poles in $\ep$ 
and the heavy-quark mass $m$ screens some of the collinear singularities
giving rise to logarithms in $m$, see~\cite{Mitov:2006xs,Catani:2000ef}. 
Recent progress on the mass dependence in the relevant soft anomalous
dimensions has been achieved in Refs.~\cite{Mitov:2009sv,Becher:2009kw,Kidonakis:2009ev}.

The common physical origin of the singularities lends itself to a
proportionality between ${\cal M}^{(m)}$ and ${\cal M}^{(m=0)}$ 
so that QCD factorization provides 
the remarkably simple and suggestive direct relation~\cite{Mitov:2006xs}
\begin{eqnarray}
  \label{eq:Mm-M0}
  {\cal M}^{(m)} &=&
  \prod_{i\in\ \{{\rm all}\ {\rm legs}\}}\,
  \left(
    Z^{(m\vert0)}_{[i]}
  \right)^{1 \over 2}\,
  \times\
  {\cal M}^{(m=0)}\, .
\end{eqnarray}
The universal multiplicative factor $Z^{(m\vert 0)}$ is process independent 
and depends only on the external parton.
Then, $Z^{(m\vert 0)}$ is defined simply 
as the ratio of the on-shell heavy-parton form factor and the corresponding massless on-shell one.
E.g. for an external heavy quark $q$, we find
\begin{equation}
\label{eq:Z}
Z^{(m\vert0)}_{[q]}\left({m^2 \over \mu^2},\alpha_{s},\ep \right) 
\, = \, 
{\cal F}_1\left({Q^2\over \mu^2},{m^2\over\mu^2},\alpha_{s},\ep \right) 
\left({\cal F}_1\left({Q^2\over \mu^2},0,\alpha_{s},\ep \right)\right)^{-1}
\, ,
\end{equation}
where only the electric form factor ${\cal F}_1$ enters in the high-energy limit $Q^2 \gg m^2$, since ${\cal F}_2$
vanishes for massless quarks.
Here $\alpha_{s}$ is evaluated at the scale $\mu^2$.
The process independence of $Z^{(m\vert0)}$ is manifest, because the hard scale $Q^2$ drops out in Eq.~(\ref{eq:Z})
leaving only a function of the ratio of scales $\mu^2/m^2$.
The definition~(\ref{eq:Z}) of  $Z^{(m\vert0)}$, however, 
excludes terms with explicit dependence on the number of heavy quarks 
(e.g. the virtual heavy loop in diagram $QL$ in Fig.~\ref{fig:massiveFF2}). 
At two-loops (and beyond) one needs additional process dependent terms for their description~\cite{Becher:2007cu}.

Applications of the formalism so far include the computation of the (virtual) 
two-loop QCD corrections to hadronic top-quark pair-production~\cite{Czakon:2007ej,Czakon:2007wk} 
in the limit when all Mandelstam invariants are large, $s, |t|, |u| \gg m^2$. 
In the same approximation, the derivation of the two-loop QED corrections to Bhabha scattering 
for a small electron mass $m_e$ has been performed~\cite{Becher:2007cu} 
(see also~\cite{Glover:2001ev,Penin:2005kf,Penin:2005eh} for earlier work).
These applications have also covered the complete dependence on virtual
contributions for heavy fermions in loops, mostly by means of 
direct calculation~\cite{Bonciani:2007eh,Actis:2007fs,Bonciani:2008ep,Actis:2008br,Kuhn:2008zs}.

With the help of Eqs.~(\ref{eq:F1rl3-x0}), (\ref{eq:Z}) 
and the known results~\cite{Moch:2005id} of the poles for the massless quark form factor 
we can extend the perturbative predictions for $Z^{(m\vert0)}$ to higher orders in $\alpha_{s}$.
Defining 
\begin{equation}
  \label{eq:Zdef-exp}
  Z^{(m\vert0)}_{[q]} 
  \, = \, 
  1 + \sum\limits_{j=1} \left({\alpha_{s} \over 4 \pi} \right)^j\, Z_{[q]}^{(j)} 
  \, \equiv \, 
  1 + \sum\limits_{j=1}\, a_s^j\, Z_{[q]}^{(j)} 
  \, ,
\end{equation}
we can derive the three-loop coefficients $Z_{[q]}^{(3)}$ up to the single poles in $\epsilon$. 

Keeping the currently unknown function $K^{(3)}$, 
see the exponential in Eq.~(\ref{eq:massFFexp}), we obtain 
{\small
\begin{eqnarray}
  \label{eq:Z1}
  Z_{[q]}^{(1)} &=& 
\cf \* \biggl\{
  {2\over \ep^2}
+ {2\*L_{\mu} + 1 \over \ep}
+ L_{\mu}^2+L_{\mu}
+ 4
+ \z2
+ \ep \* \biggl(
  {L_{\mu}^3\over 3}
+ {L_{\mu}^2\over 2}
+ (4+\z2)\*L_{\mu}
+ 8
+ {\z2 \over 2}
- {2\over 3}\*\z3
  \biggr)
+ \ep^2 \* \biggl(
  {L_{\mu}^4 \over 12}
+ {L_{\mu}^3 \over 6}
\\
& &\mbox{}
+ \biggl(
  2
  + {\z2 \over 2}
  \biggr) \* L_{\mu}^2
+ \biggl(
  8
  + {\z2 \over 2}
- {2\over 3}\*\z3
  \biggr) \* L_{\mu}
+ 16
+ 2 \* \z2
- {\z3\over 3}
+ {9\over 20} \* \z2^2
  \biggr)
+ \ep^3 \* \biggl(
  {L_{\mu}^5 \over 60}
+ {L_{\mu}^4 \over 24}
+ \biggl(
  {2\over 3}
+ {\z2 \over 6}
  \biggr)\*L_{\mu}^3
\nonumber\\
& &\mbox{}
+ \biggl(
  4
+ {\z2 \over 4}
- {\z3 \over 3}
  \biggr)\*L_{\mu}^2
+ \biggl(
  16
+ 2\*\z2
- {\z3 \over 3}
+ {9\over 20}\*\z2^2
  \biggr)\*L_{\mu}
+ 32
+ 4\*\z2
- {4\over 3}\*\z3
+ {9\over 40}\*\z2^2
- {\z2\*\z3 \over 3}
- {2\over 5}\*\z5
  \biggr)
\biggr\}
+ \calO(\ep^4)
\, ,
\nonumber\\
  \label{eq:Z2}
  Z_{[q]}^{(2)} &=& 
  \cf^2\*{2\over \ep^4}
+ {1\over \ep^3}\*\biggl\{
  \cf^2\*(4\*L_{\mu}+2)
- {11\over 2}\*\cf\*\ca
+ \nf\*\cf\biggr\}
+ {1\over\ep^2}\*\biggl\{
    \cf^2\*\biggl(4\*L_{\mu}^2+4\*L_{\mu}+{17\over 2}+2\*\z2\biggr)
\\
& &\mbox{}
  + \cf\*\ca\*\biggl(-{11\over 3}\*L_{\mu} + {17\over 9}-\z2\biggr)
  + \nf\*\cf\*\biggl({2\over 3}\*L_{\mu}-{2\over 9}\biggr)
\biggr\}
+{1\over\ep}\*\biggl\{
  \cf^2\*\biggl({8\over 3}\*L_{\mu}^3+4\*L_{\mu}^2+(17+4\*\z2)\*L_{\mu}
    +{83\over 4}
\nonumber\\
& &\mbox{}
-4\*\z2+{32\over 3}\*\z3\biggr)
+ \cf\*\ca\*\biggl(\biggl({67\over 9}-2\*\z2\biggr)\*L_{\mu}
  +{373\over 108}+{15\over 2}\*\z2-15\*\z3\biggr)
- \nf\*\cf\*\biggl({10\over 9}\*L_{\mu}+{5\over 54}+\z2\biggr)
\biggr\}
+ \cf^2\*\biggl( {4\over 3}\*L_{\mu}^4
\nonumber\\
& &\mbox{}
+{8\over 3}\*L_{\mu}^3+(17+4\*\z2)\*L_{\mu}^2
    +\biggl({83\over 2}-8\*\z2+{64\over 3}\*\z3\biggr)\*L_{\mu}
    +{561 \over 8}
    +{61\over 2}\*\z2
    -{22\over 3}\*\z3
    - 48\*\lntwo\*\z2
    -{77\over 5}\*\z2^2
    \biggr)
\nonumber\\
& &\mbox{}
+ \cf\*\ca\*\biggl(
     {11\over 9}\*L_{\mu}^3+\biggl({167\over 18}-2\*\z2\biggr)\*L_{\mu}^2
    +\biggl({1165\over 54}+{56\over 3}\*\z2-30\*\z3\biggr)\*L_{\mu}
    +{12877\over 648}
    +{323\over 18}\*\z2
    +{89\over 9}\*\z3
    +24\*\lntwo\*\z2
\nonumber\\
& &\mbox{}
    -{47\over 5}\*\z2^2
    \biggr)
+ \nf\*\cf\*\biggl(-{2\over 9}\*L_{\mu}^3-{13\over 9}\*L_{\mu}^2
    -\biggl({77\over 27}+{8\over 3}\*\z2\biggr)\*L_{\mu}
    -{1541\over 324}
    -{37\over 9}\*\z2
    -{26\over 9}\*\z3
    \biggr)
+\ep\*\biggl\{
  \cf^2\*\biggl(
  {8\over 15}\* L_{\mu}^5
+ {4\over 3}\* L_{\mu}^4
\nonumber\\
& &\mbox{}
+ \biggl(
  {34\over 3}
+ {8\over 3}\*\z2
  \biggr)\* L_{\mu}^3 
+ \biggl(
  {83\over 2}
- 8\*\z2
+ {64\over 3}\*\z3
  \biggr)\* L_{\mu}^2
+ \biggl(
  {561\over 4}
+ 61\*\z2
- {44\over 3}\*\z3
- 96\*\z2\*\lntwo
- {154\over 5}\*\z2^2
  \biggr)\* L_{\mu}
+ {723\over 16}
\nonumber\\
& &\mbox{}
+ {439\over 4}\*\z2
+ {277\over 3}\*\z3
- 24\*\z2\*\lntwo
- {677\over 5}\*\z2^2
+ 8\*\lntwof
+ 192\*\a4
+ 96\*\z2\*\lntwos
- {148\over 3}\*\z2\*\z3
- {34\over 5}\*\z5
  \biggr)
\nonumber\\
& &\mbox{}
+ \cf\*\ca\*\biggl(
  {11\over 12}\* L_{\mu}^4 
+ \biggl(
  {367\over 54}
- {4\over 3}\*\z2
  \biggr)\* L_{\mu}^3
+ \biggl(
  {1561\over 54}
+ {41\over 2}\*\z2
- 30\*\z3
  \biggr)\* L_{\mu}^2
+ \biggl(
  {22381\over 324}
+ {679\over 18}\*\z2
+ {52\over 3}\*\z3
+ 48\*\z2\*\lntwo
\nonumber\\
& &\mbox{}
- {94\over 5}\*\z2^2
  \biggr)\* L_{\mu}
+ {629821\over 3888}
+ {7805\over 108}\*\z2
- {1162\over 27}\*\z3
+ 12\*\z2\*\lntwo
+ {209\over 2}\*\z2^2
- 48\*\z2\*\lntwos
- 4\*\lntwof
- 96\*\a4
\nonumber\\
& &\mbox{}
+ {83\over 3}\*\z2\*\z3
- 208\*\z5
\biggr)
+ \nf\*\cf\*\biggl(
- {1\over 6}\* L_{\mu}^4 
- {29\over 27}\* L_{\mu}^3
- \biggl(
  {113\over 27}
+ 3\*\z2
  \biggr)\* L_{\mu}^2
- \biggl(
  {2405\over 162}
+ {77\over 9}\*\z2
+ {16\over 3}\*\z3
  \biggr)\* L_{\mu}
\nonumber\\
& &\mbox{}
- {46205\over 1944}
- {673\over 54}\*\z2
- {182\over 27}\*\z3
- {49\over 5}\*\z2^2
\biggr)
\biggr\}
+ \calO(\ep^2)
\nonumber\\
  \label{eq:Z3}
  Z_{[q]}^{(3)} &=& 
  \cf^3\*{4\over 3}\* {1\over \ep^6}
+ {1\over \ep^5}\*\biggl\{
  \cf^3\*(4\*L_{\mu}+2)
- 11\*\cf^2\*\ca
+ 2\*\nf\*\cf^2\biggr\}
+ {1\over \ep^4}\*\biggl\{
  \cf^3\*\biggl(
  6\*L_{\mu}^2 
+ 6\*L_{\mu} 
+ 9
+ 2\*\z2
\biggr)
\\
& &\mbox{}
- \cf^2\*\ca\*\biggl(
  {55\over 3}\*L_{\mu}
+ {31\over 18}
+ 2\*\z2
\biggr)
+ {1331\over 81}\*\cf\*\ca^2
+ \nf\*\cf^2\*\biggl(
  {10\over 3}\*L_{\mu}
+ {5\over 9}
\biggr)
- {484\over 81}\*\nf\*\cf\*\ca
+ {44\over 81}\*\nf^2\*\cf
\biggr\}
\nonumber\\
& &\mbox{}
+ {1\over \ep^3}\*\biggl\{
  \cf^3\*\biggl(
  6\*L_{\mu}^3 
+ 9\*L_{\mu}^2
+ \biggl(
  27
+ 6\*\z2
  \biggr)\* L_{\mu}
+ {77\over 3}
+ {68\over 3}\*\z3
- 9\*\z2
\biggr)
- \cf^2\*\ca\*\biggl(
  {77\over 6}\*L_{\mu}^2
- \biggl(
  {19\over 2}
- 6\*\z2\biggr)\*L_{\mu}
+ {713\over 54}
\nonumber\\
& &\mbox{}
+ 30\*\z3
- {17\over 2}\*\z2
\biggr)
+ \cf\*\ca^2\*\biggl(
  {242\over 27}\*L_{\mu}
- {5044\over 243}
+ {110\over 27}\*\z2
\biggr)
+ \nf\*\cf^2\*\biggl(
  {7\over 3}\*L_{\mu}^2
- L_{\mu}
+ {145\over 27}
- \z2
\biggr)
\nonumber\\
& &\mbox{}
- \nf\*\cf\*\ca\*\biggl(
  {88\over 27}\*L_{\mu}
- {1544\over 243}
+ {20\over 27}\*\z2
\biggr)
+ \nf^2\*\cf\*\biggl(
  {8\over 27}\*L_{\mu}
- {64\over 243}
\biggr)
\biggr\}
+ {1\over \ep^2}\*\biggl\{
  \cf^3\*\biggl(
  {9\over 2}\*L_{\mu}^4 
+ 9\*L_{\mu}^3
+ \biggl(
  {81\over 2}
+ 9\*\z2
  \biggr)\*L_{\mu}^2
\nonumber\\
& &\mbox{}
+ \biggl(
  77
- 27\*\z2
+ 68\*\z3
  \biggr)\*L_{\mu}
+ 111
+ {103\over 2}\*\z2
- 2\*\z3
- 96\*\z2\*\lntwo
- {317\over 10}\*\z2^2
\biggr)
+ \cf^2\*\ca\*\biggl(
- {55\over 18}\*L_{\mu}^3
\nonumber\\
& &\mbox{}
+ \biggl(
  {347\over 12}
- 9\*\z2
\biggr) \* L_{\mu}^2
+ \biggl(
  {409\over 18}
+ {241\over 6}\*\z2
- 90\*\z3
\biggr) \* L_{\mu}
+ {797\over 162}
+ {1915\over 36}\*\z2
- {188\over 9}\*\z3
+ 48\*\z2\*\lntwo
- {99\over 5}\*\z2^2
\biggr)
\nonumber\\
& &\mbox{}
+ \cf\*\ca^2\*\biggl(
  \biggl(
- {2086\over 81}
+ {44\over 9}\*\z2
  \biggr)\*L_{\mu}          
- {1529\over 486}
+ {1034\over 27}\*\z3
- {2021\over 81}\*\z2
+ {88\over 45}\*\z2^2
\biggr)
+ \nf\*\cf^2\*\biggl(
  {5\over 9} \* L_{\mu}^3
- {25\over 6} \* L_{\mu}^2 
\nonumber\\
& &\mbox{}
+ \biggl(
  {7\over 9}
- {17\over 3}\*\z2
  \biggr)\*L_{\mu}          
- {286\over 81}
- {209\over 18}\*\z2
+ {2\over 3}\*\z3
\biggr)
+ \nf\*\cf\*\ca \*\biggl(
  \biggl(
    {668 \over 81}
  - {8 \over 9}\*\z2
  \biggr)\*L_{\mu} 
+ {280 \over 243}
- {236 \over 27}\*\z3
+ {548 \over 81}\*\z2
\biggr)
\nonumber\\
& &\mbox{}
- \nf^2\*\cf\*\biggl(
  {40\over 81}\*L_{\mu}
+ {2\over 27}
+ {4\over 9}\*\z2
\biggr)
\biggr\}
+ {1\over \ep}\*\biggl\{
  \cf^3 \*\biggl(
  {27 \over 10}\*L_{\mu}^5
+ {27 \over 4}\*L_{\mu}^4
+ \biggl(
  {81 \over 2}
+ 9 \* \z2
\biggr)\*L_{\mu}^3
+ \biggl(
  {231 \over 2}
- {81 \over 2} \* \z2
+ 102 \* \z3
\biggr)\*L_{\mu}^2 
\nonumber\\
& &\mbox{}
+ \biggl(
  333
+ {309 \over 2} \* \z2
- 288 \* \z2 \* \lntwo
- 6 \* \z3
- {951 \over 10} \* \z2^2
\biggr)\*L_{\mu} 
+ {127 \over 2}
+ {429 \over 2} \* \z2
+ 229 \* \z3
- 96 \* \z2 \* \lntwo
+ 192 \* \z2 \* \lntwos
\nonumber\\
& &\mbox{}
+ 16 \* \lntwof
+ 384 \* \a4
- {5871 \over 20} \* \z2^2
- 86 \* \z2 \* \z3
- {64 \over 5} \* \z5
\biggr)
+ \cf^2 \* \ca \*\biggl(
  {187 \over 72} \* L_{\mu}^4
+ \biggl(
  {1393 \over 36}
- 9 \* \z2
\biggr)\*L_{\mu}^3
\nonumber\\
& &\mbox{}
+ \biggl(
  {1157 \over 12}
+ {899 \over 12} \* \z2
- 135 \* \z3
\biggr)\*L_{\mu}^2 
+ \biggl(
  {4507 \over 27}
+ {521 \over 4} \* \z2
+ 144 \* \z2 \* \lntwo
+ {140 \over 9} \* \z3
- {297 \over 5} \* \z2^2
\biggr)\*L_{\mu} 
+ {138403 \over 486}
\nonumber\\
& &\mbox{}
+ {19211 \over 108} \* \z2
+ 48 \* \z2 \* \lntwo
- {2441 \over 18} \* \z3
- 96 \* \z2 \* \lntwos
+ {1633 \over 8} \* \z2^2
- 8 \* \lntwof
- 192 \* \a4
+ 41 \* \z2 \* \z3
- 416 \* \z5
\biggr)
\nonumber\\
& &\mbox{}
+ \cf \* \ca^2 \*\biggl(
  {245 \over 9}
- {536 \over 27} \* \z2
+ {44 \over 9} \* \z3
+ {88 \over 15} \* \z2^2
\biggr)\*L_{\mu} 
+ \nf\*\cf^2 \*\biggl(
- {107 \over 18}\*L_{\mu}^3
- {17 \over 36}\*L_{\mu}^4
- \biggl(
  {73 \over 6}
+ {67 \over 6} \* \z2
  \biggr)\*L_{\mu}^2
\nonumber\\
& &\mbox{}
- \biggl(
  {1087 \over 27}
+ {59 \over 2} \* \z2
+ {110 \over 9} \* \z3
  \biggr)\*L_{\mu} 
- {9341 \over 243}
- {1687 \over 54} \* \z2
- {149 \over 9} \* \z3
- {403 \over 20} \* \z2^2
\biggr)
\nonumber\\
& &\mbox{}
- \nf\*\cf\*\ca \*\biggl(
  {418 \over 81}
- {80 \over 27} \* \z2
+ {56 \over 9} \* \z3
\biggr)\*L_{\mu} 
- \nf^2\*\cf \* {8 \over 81} \* L_{\mu} 
- {1 \over 6} \* K^{(3)}
\biggr\}
+ \calO(\ep^0)
\nonumber
\, ,
\end{eqnarray}
}
where $L_{\mu}$ is defined in Eq.~(\ref{eq:Lmu-def}).
Eq.~(\ref{eq:Z1}) has been obtained from an all-order in $\epsilon$ formula 
in Ref.~\cite{Mitov:2006xs} based on $Z_{[q]}^{(1)}$ being identical 
to the virtual contribution (prior to collinear factorization) 
to the perturbative fragmentation function of a heavy quark~\cite{Melnikov:2004bm}.

According to the discussion above all contributions arising from virtual heavy-flavor 
loops are omitted here and it requires substantially more effort to 
determine them in massive scattering amplitudes,
see e.g. Refs.~\cite{Czakon:2007ej,Czakon:2007wk} for the two-loop 
QCD corrections to hadronic top-quark pair-production.
It should be clear though, that the highest power in $\nf = \nl + \nh$ 
is always governed by the renormalization of the coupling constant $\alpha_s$. 
Thus, terms proportional to the maximal power $\nh$ 
can be predicted on the basis of the $Z^{(m\vert0)}$-factor 
for massive scattering amplitudes as well.
However, sub-leading terms in $\nh$ remain inaccessible or require extending the results of Ref.~\cite{Becher:2007cu}.

With the recent progress in the computation of the massless quark form factor~\cite{Baikov:2009bg,Heinrich:2009be,Toedtli:2009bn},
one can hope to extend Eq.~(\ref{eq:Z3}) to the constant terms as well.
However, to that end, we would need to know also the high energy-limit of ${\cal F}_1^{(2)}$ even to order ${\cal O}(\epsilon^2)$
as well as that of ${\cal F}_1^{(3)}$ to order ${\cal O}(\epsilon^0)$.
In terms of the functions $K$ and $C$ governing the exponentiation~(\ref{eq:massFFexp}) 
this requires besides $K^{(3)}$ also the two- and three-loop matching
coefficients $C^{(2)}$ and $C^{(3)}$ to sufficient depth in $\epsilon$.
Of these quantities, at least the coefficient $C^{(2)}$ is not
entirely out of reach, because the necessary two-loop master integrals are known  
to sufficient depth in $\ep$ (see Sec.~\ref{sec:calculation}).

In closing, we comment once more on the importance of Eqs.~(\ref{eq:Mm-M0}) and (\ref{eq:Z1})--(\ref{eq:Z3}).
We can derive all logarithmically enhanced terms and even the constant, i.e. mass-independent terms of ${\cal M}^{(m)}$,
provided ${\cal M}^{(m=0)}$ and all $Z_{[q]}^{(i)}$ are known to sufficient depth in $\epsilon$. 
For massless scattering amplitudes we do understand the general structure of singularities~\cite{Catani:1998bh,Sterman:2002qn}.
This allows to predict the pole structure in $\ep$ of massless amplitudes 
at any order based on a small number of perturbatively calculable anomalous dimensions
(see also~\cite{Dixon:2008gr,Becher:2009cu,Gardi:2009qi,Dixon:2009gx}).
This is a constructive approach to ${\cal M}^{(m=0)}$. 
Thus, with the help of Eq.~(\ref{eq:Mm-M0}), the singular limit of ${\cal M}^{(m=0)}$ 
and the factors $Z^{(i)}$, $i = 1, \dots, 3$ from Eqs.~(\ref{eq:Z1})--(\ref{eq:Z3}) 
one can study the logarithmically enhanced terms in an amplitude ${\cal M}^{(m)}$ 
with external massive fermions.
It is now, for instance, a straightforward exercise to investigate explicit 
results for the virtual contributions to massive-fermion scattering amplitudes in the
small-mass limit at three loops, 
an obvious example being e.g. the three-loop virtual QED amplitude of Bhabha scattering 
to logarithmic accuracy.
We leave these issues to future work.

%
%
%
\section{Conclusions}
\label{sec:conclusion}
%
%

The form factor of heavy quarks has been studied. 
Its radiative corrections in QCD are of great relevance for precision predictions of many observables at colliders.
Moreover, on the formal side, they exhibit very interesting structures of quantum field theory.

We have presented a fully rigorous calculation of the two-loop QCD corrections
to the electric and magnetic form factors ${\cal F}_1$ and ${\cal F}_2$ in dimensional regularization.
This allows for an independent check of the previous computation reported 
in the literature~\cite{Bernreuther:2004ih}.
In addition we have obtained new two-loop results at order $\epsilon$.
To that end, we have determined the Laurent series of the necessary master integrals 
to very high order in $\epsilon$.
The expansion of our result in various limits has delivered new insight into
the anatomy of the two-loop form factor near threshold 
and the Coulomb singularities for color-singlet and  octet currents.
The high-energy limit on the other hand provided new tests of the exponentiation in $D$ dimensions
as well as an improved three-loop prediction for the form factor 
to logarithmic accuracy in the heavy-quark mass $m$.
We have also discussed the consequences for massive scattering amplitudes, 
where we have been able to advance towards three-loop predictions for massive $n$-parton 
amplitudes in the small-mass limit based on the known singularity structure of the massless result.
As the heavy-quark form factor and its QCD corrections are essential ingredients in high precision
theory predictions, we believe, there will be several other future applications, 
where our results will prove to be useful.

Files of our results can be obtained from the preprint server 
{\tt http://arXiv.org} by downloading the source. 
Furthermore they are available at~\cite{Zeuthen-CAS:2009} 
or from the authors upon request.

\bigskip

{\bf{Acknowledgments:}}
We would like to thank P.~Mastrolia for assistance 
in the comparison to the results of Ref.~\cite{Bernreuther:2004ih} 
and M.~Kalmykov for explanations concerning the program {\ttfamily ON-SHELL2}~\cite{Fleischer:1999tu}.
We are also grateful to M.~Steinhauser for providing Eq.~(\ref{eq:alphas-dec-ep}).
This work is supported in part by the Deutsche Forschungsgemeinschaft in Sonderforschungs\-be\-reich/Transregio~9 
and by the European Community in Marie-Curie Research Training Networks MRTN-CT-2006-035505 ``HEPTOOLS'' 
and MRTN-CT-2006-035482 ``FLAVIAnet''.
A.M. is supported by a fellowship from the US LHC Theory Initiative through NSF grant 0653342 
and S.M. acknowledges the Helmholtz Gemeinschaft under contract VH-NG-105.

%
%
{\footnotesize


\begingroup\begin{thebibliography}{10}

\bibitem{Bernreuther:2004ih}
W.~Bernreuther, R.~Bonciani, T.~Gehrmann, R.~Heinesch, T.~Leineweber,
  P.~Mastrolia, and E.~Remiddi, { Nucl. Phys.} { B706} (2005) 245--324,
\href{http://xxx.lanl.gov/abs/hep-ph/0406046}{{\tt hep-ph/0406046}}.

\bibitem{Bernreuther:2004th}
W.~Bernreuther, R.~Bonciani, T.~Gehrmann, R.~Heinesch, T.~Leineweber,
  P.~Mastrolia, and E.~Remiddi, { Nucl. Phys.} { B712} (2005) 229--286,
\href{http://xxx.lanl.gov/abs/hep-ph/0412259}{{\tt hep-ph/0412259}}.

\bibitem{Bernreuther:2005rw}
W.~Bernreuther, R.~Bonciani, T.~Gehrmann, R.~Heinesch, T.~Leineweber,
  and E.~Remiddi, { Nucl. Phys.} { B723} (2005) 91--116,
\href{http://xxx.lanl.gov/abs/hep-ph/0504190}{{\tt hep-ph/0504190}}.

\bibitem{Sudakov:1954sw}
V.~V. Sudakov, { Sov. Phys. JETP} { 3} (1956)
65--71.

\bibitem{Collins:1980ih}
J.~C. Collins, { Phys. Rev.} { D22} (1980)
1478.

\bibitem{Mitov:2006xs}
A.~Mitov and S.~Moch, { JHEP} { 05} (2007) 001,
\href{http://xxx.lanl.gov/abs/hep-ph/0612149}{{\tt hep-ph/0612149}}.

\bibitem{Catani:2000ef}
S.~Catani, S.~Dittmaier, and Z.~Trocsanyi, { Phys. Lett.} { B500} (2001)
  149--160,
\href{http://xxx.lanl.gov/abs/hep-ph/0011222}{{\tt hep-ph/0011222}}.

\bibitem{Becher:2007cu}
T.~Becher and K.~Melnikov, { JHEP} { 06} (2007) 084,
\href{http://xxx.lanl.gov/abs/arXiv:0704.3582}{{\tt arXiv:0704.3582}}.

\bibitem{Catani:1998bh}
S.~Catani, { Phys. Lett.} { B427} (1998) 161--171,
\href{http://xxx.lanl.gov/abs/hep-ph/9802439}{{\tt hep-ph/9802439}}.

\bibitem{Sterman:2002qn}
G.~Sterman and M.~E. Tejeda-Yeomans, { Phys. Lett.} { B552} (2003) 48--56,
\href{http://xxx.lanl.gov/abs/hep-ph/0210130}{{\tt hep-ph/0210130}}.

\bibitem{Czakon:2007ej}
M.~Czakon, A.~Mitov, and S.~Moch, { Phys. Lett.} { B651} (2007) 147--159,
\href{http://xxx.lanl.gov/abs/arXiv:0705.1975}{{\tt arXiv:0705.1975}}.

\bibitem{Czakon:2007wk}
M.~Czakon, A.~Mitov, and S.~Moch, { Nucl. Phys.} { B798} (2008) 210--250,
\href{http://xxx.lanl.gov/abs/arXiv:0707.4139}{{\tt arXiv:0707.4139}}.

\bibitem{Mitov:2009sv}
A.~Mitov, G.~Sterman, and I.~Sung,
\href{http://xxx.lanl.gov/abs/arXiv:0903.3241}{{\tt arXiv:0903.3241}}.

\bibitem{Becher:2009kw}
T.~Becher and M.~Neubert,
\href{http://xxx.lanl.gov/abs/arXiv:0904.1021}{{\tt arXiv:0904.1021}}.

\bibitem{Moch:2005id}
S.~Moch, J.~A.~M. Vermaseren, and A.~Vogt, { JHEP} { 08} (2005) 049,
\href{http://xxx.lanl.gov/abs/hep-ph/0507039}{{\tt hep-ph/0507039}}.

\bibitem{Laporta:1996mq}
S.~Laporta and E.~Remiddi, { Phys. Lett.} { B379} (1996) 283--291,
\href{http://xxx.lanl.gov/abs/hep-ph/9602417}{{\tt hep-ph/9602417}}.

\bibitem{Laporta:2001dd}
S.~Laporta, { Int. J. Mod. Phys.} { A15} (2000) 5087--5159,
\href{http://xxx.lanl.gov/abs/hep-ph/0102033}{{\tt hep-ph/0102033}}.

\bibitem{Anastasiou:2004vj}
C.~Anastasiou and A.~Lazopoulos, { JHEP} { 07} (2004) 046,
\href{http://xxx.lanl.gov/abs/hep-ph/0404258}{{\tt hep-ph/0404258}}.

\bibitem{Czakon:2004tg}
M.~Czakon, J.~Gluza, and T.~Riemann, { Nucl. Phys. Proc. Suppl.} { 135} (2004)
  83--87,
\href{http://xxx.lanl.gov/abs/hep-ph/0406203}{{\tt hep-ph/0406203}}.

\bibitem{Czakon:2004wm}
M.~Czakon, J.~Gluza, and T.~Riemann, { Phys. Rev.} { D71} (2005) 073009,
\href{http://xxx.lanl.gov/abs/hep-ph/0412164}{{\tt hep-ph/0412164}}.

\bibitem{Fleischer:1999tu}
J.~Fleischer and M.~Kalmykov, { Comput. Phys. Commun.} { 128} (2000) 531--549,
\href{http://xxx.lanl.gov/abs/hep-ph/9907431}{{\tt hep-ph/9907431}}.

\bibitem{Remiddi:1999ew}
E.~Remiddi and J.~A.~M. Vermaseren, { Int. J. Mod. Phys.} { A15} (2000) 725,
\href{http://xxx.lanl.gov/abs/hep-ph/9905237}{{\tt hep-ph/9905237}}.

\bibitem{'tHooft:1979xw}
G.~'t~Hooft and M.~Veltman, { Nucl. Phys.} { B153} (1979)
365--401.

\bibitem{Fleischer:1999hp}
J.~Fleischer, M.~Kalmykov, and A.~Kotikov, { Phys. Lett.} { B462} (1999)
  169--177,
\href{http://xxx.lanl.gov/abs/hep-ph/9905249}{{\tt hep-ph/9905249}}.

\bibitem{Davydychev:2003mv}
A.~Davydychev and M.~Kalmykov, { Nucl. Phys.} { B699} (2004) 3--64,
\href{http://xxx.lanl.gov/abs/hep-th/0303162}{{\tt hep-th/0303162}}.

\bibitem{Broadhurst:1990ei}
D.~Broadhurst, { Z. Phys.} { C47} (1990)
115--124.

\bibitem{Broadhurst:1991fi}
D.~Broadhurst, { Z. Phys.} { C54} (1992)
599--606.

\bibitem{Bonciani:2003te}
R.~Bonciani, P.~Mastrolia, and E.~Remiddi, { Nucl. Phys.} { B661} (2003)
  289--343,
\href{http://xxx.lanl.gov/abs/hep-ph/0301170}{{\tt hep-ph/0301170}}.

\bibitem{Kotikov:1991hm}
A.~Kotikov, { Phys. Lett.} { B259} (1991)
314--322.

\bibitem{Remiddi:1997ny}
E.~Remiddi, { Nuovo Cim.} { A110} (1997) 1435--1452,
\href{http://xxx.lanl.gov/abs/hep-th/9711188}{{\tt hep-th/9711188}}.

\bibitem{Zeuthen-CAS:2009}
DESY, webpage
  http://www-zeuthen.desy.de/theory/research/\href{http://www-zeuthen.desy.de/%
theory/research/CAS.html}{CAS.html}.

\bibitem{Maitre:2005uu}
D.~Maitre, { Comput. Phys. Commun.} { 174} (2006) 222--240,
\href{http://xxx.lanl.gov/abs/hep-ph/0507152}{{\tt hep-ph/0507152}}.

\bibitem{Bogner:2007cr}
C.~Bogner and S.~Weinzierl, { Comput. Phys. Commun.} { 178} (2008) 596--610,
\href{http://xxx.lanl.gov/abs/arXiv:0709.4092}{{\tt arXiv:0709.4092}}.

\bibitem{Gluza:2007rt}
J.~Gluza, K.~Kajda, and T.~Riemann, { Comput. Phys. Commun.} { 177} (2007)
  879--893,
\href{http://xxx.lanl.gov/abs/arXiv:0704.2423}{{\tt arXiv:0704.2423}}.

\bibitem{Czakon:2005rk}
M.~Czakon, { Comput. Phys. Commun.} { 175} (2006) 559--571,
\href{http://xxx.lanl.gov/abs/hep-ph/0511200}{{\tt hep-ph/0511200}}.

\bibitem{Melnikov:2000qh}
K.~Melnikov and T.~van Ritbergen, { Phys. Lett.} { B482} (2000) 99--108,
\href{http://xxx.lanl.gov/abs/hep-ph/9912391}{{\tt hep-ph/9912391}}.

\bibitem{Melnikov:2000zc}
K.~Melnikov and T.~van Ritbergen, { Nucl. Phys.} { B591} (2000) 515--546,
\href{http://xxx.lanl.gov/abs/hep-ph/0005131}{{\tt hep-ph/0005131}}.

\bibitem{Vermaseren:2000nd}
J.~A.~M. Vermaseren,
\href{http://xxx.lanl.gov/abs/math-ph/0010025}{{\tt math-ph/0010025}}.

\bibitem{Czarnecki:1997vz}
A.~Czarnecki and K.~Melnikov, { Phys. Rev. lett.} { 80} (1998) 2531--2534,
\href{http://xxx.lanl.gov/abs/hep-ph/9712222}{{\tt hep-ph/9712222}}.

\bibitem{Pineda:2006ri}
A.~Pineda and A.~Signer, { Nucl. Phys.} { B762} (2007) 67--94,
\href{http://xxx.lanl.gov/abs/hep-ph/0607239}{{\tt hep-ph/0607239}}.

\bibitem{Petrelli:1997ge}
A.~Petrelli, M.~Cacciari, M.~Greco, F.~Maltoni, and M.~L. Mangano, { Nucl.
  Phys.} { B514} (1998) 245--309,
\href{http://xxx.lanl.gov/abs/hep-ph/9707223}{{\tt hep-ph/9707223}}.

\bibitem{Hagiwara:2008df}
K.~Hagiwara, Y.~Sumino, and H.~Yokoya, { Phys. Lett.} { B666} (2008) 71--76,
\href{http://xxx.lanl.gov/abs/arXiv:0804.1014}{{\tt arXiv:0804.1014}}.

\bibitem{Kiyo:2008bv}
Y.~Kiyo, J.~H. K\"uhn, S.~Moch, M.~Steinhauser, and P.~Uwer,
\href{http://xxx.lanl.gov/abs/arXiv:0812.0919}{{\tt arXiv:0812.0919}}.

\bibitem{Collins:1989bt}
J.~C. Collins, { Adv. Ser. Direct. High Energy Phys.} { 5} (1989) 573--614,
\href{http://xxx.lanl.gov/abs/hep-ph/0312336}{{\tt hep-ph/0312336}}.

\bibitem{Kodaira:1982nh}
J.~Kodaira and L.~Trentadue, { Phys. Lett.} { 112B} (1982)
66.

\bibitem{Moch:2004pa}
S.~Moch, J.~A.~M. Vermaseren, and A.~Vogt, { Nucl. Phys.} { B688} (2004)
  101--134,
\href{http://xxx.lanl.gov/abs/hep-ph/0403192}{{\tt hep-ph/0403192}}.

\bibitem{Vogt:2004mw}
A.~Vogt, S.~Moch, and J.~A.~M. Vermaseren, { Nucl. Phys.} { B691} (2004)
  129--181,
\href{http://xxx.lanl.gov/abs/hep-ph/0404111}{{\tt hep-ph/0404111}}.

\bibitem{Magnea:1990zb}
L.~Magnea and G.~Sterman, { Phys. Rev.} { D 42} (1990) 4222.

\bibitem{Moch:2005tm}
S.~Moch, J.~A.~M. Vermaseren, and A.~Vogt, { Phys. Lett.} { B625} (2005)
  245--252,
\href{http://xxx.lanl.gov/abs/hep-ph/0508055}{{\tt hep-ph/0508055}}.

\bibitem{Dixon:2008gr}
L.~J. Dixon, L.~Magnea, and G.~Sterman, { JHEP} { 08} (2008) 022,
\href{http://xxx.lanl.gov/abs/0805.3515}{{\tt 0805.3515}}.

\bibitem{Kuhn:2001hz}
J.~H. K\"uhn, S.~Moch, A.~A. Penin, and V.~A. Smirnov, { Nucl. Phys.} { B616}
  (2001) 286--306,
\href{http://xxx.lanl.gov/abs/hep-ph/0106298}{{\tt hep-ph/0106298}}.

\bibitem{Feucht:2003yx}
B.~Feucht, J.~H. K\"uhn, and S.~Moch, { Phys. Lett.} { B561} (2003) 111--118,
\href{http://xxx.lanl.gov/abs/hep-ph/0303016}{{\tt hep-ph/0303016}}.

\bibitem{Appelquist:1974tg}
T.~Appelquist and J.~Carazzone, { Phys. Rev.} { D11} (1975)
2856.

\bibitem{Larin:1994va}
S.~A. Larin, T.~van Ritbergen, and J.~A.~M. Vermaseren, { Nucl. Phys.} { B438}
  (1995) 278--306,
\href{http://xxx.lanl.gov/abs/hep-ph/9411260}{{\tt hep-ph/9411260}}.

\bibitem{Chetyrkin:1997un}
K.~G. Chetyrkin, B.~A. Kniehl, and M.~Steinhauser, { Nucl. Phys.} { B510}
  (1998) 61--87,
\href{http://xxx.lanl.gov/abs/hep-ph/9708255}{{\tt hep-ph/9708255}}.

\bibitem{Schroder:2005hy}
Y.~Schr\"oder and M.~Steinhauser, { JHEP} { 01} (2006) 051,
\href{http://xxx.lanl.gov/abs/hep-ph/0512058}{{\tt hep-ph/0512058}}.

\bibitem{Chetyrkin:2005ia}
K.~G. Chetyrkin, J.~H. K\"uhn, and C.~Sturm, { Nucl. Phys.} { B744} (2006)
  121--135,
\href{http://xxx.lanl.gov/abs/hep-ph/0512060}{{\tt hep-ph/0512060}}.

\bibitem{Baikov:2009bg}
P.~A. Baikov, K.~G. Chetyrkin, A.~V. Smirnov, V.~A. Smirnov, and
  M.~Steinhauser,
\href{http://xxx.lanl.gov/abs/arXiv:0902.3519}{{\tt arXiv:0902.3519}}.

\bibitem{Aybat:2006wq}
S.~M. Aybat, L.~J. Dixon, and G.~Sterman, { Phys. Rev. Lett.} { 97} (2006)
  072001,
\href{http://xxx.lanl.gov/abs/hep-ph/0606254}{{\tt hep-ph/0606254}}.

\bibitem{Aybat:2006mz}
S.~M. Aybat, L.~J. Dixon, and G.~Sterman, { Phys. Rev.} { D74} (2006) 074004,
\href{http://xxx.lanl.gov/abs/hep-ph/0607309}{{\tt hep-ph/0607309}}.

\bibitem{Becher:2009cu}
T.~Becher and M.~Neubert,
\href{http://xxx.lanl.gov/abs/arXiv:0901.0722}{{\tt arXiv:0901.0722}}.

\bibitem{Gardi:2009qi}
E.~Gardi and L.~Magnea,
\href{http://xxx.lanl.gov/abs/arXiv:0901.1091}{{\tt arXiv:0901.1091}}.

\bibitem{Dixon:2009gx}
L.~J. Dixon,
\href{http://xxx.lanl.gov/abs/arXiv:0901.3414}{{\tt arXiv:0901.3414}}.

\bibitem{Kidonakis:2009ev}
N.~Kidonakis,
\href{http://xxx.lanl.gov/abs/arXiv:0903.2561}{{\tt arXiv:0903.2561}}.

\bibitem{Glover:2001ev}
E.~W.~N. Glover, J.~B. Tausk, and J.~J. Van~der Bij, { Phys. Lett.} { B516}
  (2001) 33--38,
\href{http://xxx.lanl.gov/abs/hep-ph/0106052}{{\tt hep-ph/0106052}}.

\bibitem{Penin:2005kf}
A.~A. Penin, { Phys. Rev. Lett.} { 95} (2005) 010408,
\href{http://xxx.lanl.gov/abs/hep-ph/0501120}{{\tt hep-ph/0501120}}.

\bibitem{Penin:2005eh}
A.~A. Penin, { Nucl. Phys.} { B734} (2006) 185--202,
\href{http://xxx.lanl.gov/abs/hep-ph/0508127}{{\tt hep-ph/0508127}}.

\bibitem{Bonciani:2007eh}
R.~Bonciani, A.~Ferroglia, and A.~A. Penin, { Phys. Rev. Lett.} { 100} (2008)
  131601,
\href{http://xxx.lanl.gov/abs/arXiv:0710.4775}{{\tt arXiv:0710.4775}}.

\bibitem{Actis:2007fs}
S.~Actis, M.~Czakon, J.~Gluza, and T.~Riemann, { Phys. Rev. Lett.} { 100}
  (2008) 131602,
\href{http://xxx.lanl.gov/abs/arXiv:0711.3847}{{\tt arXiv:0711.3847}}.

\bibitem{Bonciani:2008ep}
R.~Bonciani, A.~Ferroglia, and A.~A. Penin, { JHEP} { 02} (2008) 080,
\href{http://xxx.lanl.gov/abs/arXiv:0802.2215}{{\tt arXiv:0802.2215}}.

\bibitem{Actis:2008br}
S.~Actis, M.~Czakon, J.~Gluza, and T.~Riemann, { Phys. Rev.} { D78} (2008)
  085019,
\href{http://xxx.lanl.gov/abs/arXiv:0807.4691}{{\tt arXiv:0807.4691}}.

\bibitem{Kuhn:2008zs}
J.~H. K\"uhn and S.~Uccirati, { Nucl. Phys.} { B806} (2009) 300--326,
\href{http://xxx.lanl.gov/abs/arXiv:0807.1284}{{\tt arXiv:0807.1284}}.

\bibitem{Melnikov:2004bm}
K.~Melnikov and A.~Mitov, { Phys. Rev.} { D70} (2004) 034027,
\href{http://xxx.lanl.gov/abs/hep-ph/0404143}{{\tt hep-ph/0404143}}.

\bibitem{Heinrich:2009be}
G.~Heinrich, T.~Huber, D.~A. Kosower, and V.~A. Smirnov,
\href{http://xxx.lanl.gov/abs/arXiv:0902.3512}{{\tt arXiv:0902.3512}}.

\bibitem{Toedtli:2009bn}
B.~T\"odtli,
\href{http://xxx.lanl.gov/abs/arXiv:0903.0540}{{\tt arXiv:0903.0540}}.

\end{thebibliography}\endgroup

\providecommand{\href}[2]{#2}

}

\end{document}